\documentclass[aps,prl,preprintnumbers,amsmath,amssymb,latexsym,nofootinbib,array,enumerate,letter,twocolumn,superscriptaddress]{revtex4-1}
\usepackage{blindtext}
\usepackage{lineno}
\usepackage{bm,color,xcolor}
\usepackage{slashed} 
\usepackage{graphicx}
\usepackage{subfigure} 
\usepackage{amsmath}
\usepackage{amssymb}
\usepackage{multirow}
\usepackage{float}
\usepackage{comment}
\usepackage{booktabs}
\usepackage{fontawesome} \usepackage{mathtools}
\usepackage[colorlinks=true
,urlcolor=blue
,anchorcolor=blue
,citecolor=blue 
,filecolor=blue
,linkcolor=blue
,menucolor=blue
,linktocpage=true
,pdfproducer=medialab
,pdfa=true
]{hyperref}
\usepackage{lipsum}

\makeatletter
\def\l@subsubsection#1#2{}
\makeatother

\newcommand{\cm}{\ensuremath{\mathrm{cm}}}

\newcommand{\eV}{\ensuremath{\mathrm{eV}}}

\newcommand{\GeV}{\ensuremath{\mathrm{GeV}}}

\newcommand{\brr}[1]{\left(#1\right)}
\newcommand{\srr}[1]{\left[#1\right]}

\begin{document}

\title{Probing Axion via Mössbauer Spectroscopy}

\author{Shengyi Liu} 
\email{sliufp@connect.ust.hk}
\affiliation{Department of Physics and Jockey Club Institute for Advanced Study, The Hong Kong University of Science and Technology, Hong Kong S.A.R., P.R.China}
\author{Kun-Feng Lyu} 
\email{kunfeng.lyu@ou.edu}
\affiliation{Homer L. Dodge Department of Physics and Astronomy, University of Oklahoma, Norman, OK 73019, USA}
\affiliation{Department of Physics and Astronomy, University of Utah, Salt Lake City, UT 84112, USA}
\author{Jie Meng} 
\email{mengj@pku.edu.cn}
\affiliation{State Key Laboratory of Nuclear Physics and Technology, School of Physics, Peking University,
Beijing 100871, China}
\author{Jing Shu}
\email{jshu@pku.edu.cn}
\affiliation{State Key Laboratory of Nuclear Physics and Technology, School of Physics, Peking University,
Beijing 100871, China}
\affiliation{Center for High Energy Physics, Peking University, Beijing 100871, China}
\affiliation{Beijing Laser Acceleration Innovation Center, Huairou, Beijing, 101400, China}
\author{Yakun Wang} 
\email{wangyk@buaa.edu.cn}
\affiliation{School of Physics, Beihang University, Beijing 102206, China}
\author{Yue Zhao}
\email{zhaoyue@ust.hk}
\affiliation{Department of Physics and Jockey Club Institute for Advanced Study, The Hong Kong University of Science and Technology, Hong Kong S.A.R., P.R.China}

\begin{abstract}

We propose using the ultra-narrow 88 keV Mössbauer transition in $^{109}$Ag to search for QCD axion dark matter. The sub-eV axion field oscillates coherently, inducing a time-varying effective $\bar{\theta}_{\rm QCD}$ angle. This, in turn, modulates the nuclear binding energy. From existing linewidth measurements, we derive constraints on the $f^{-1}_a$–$m_a$ plane that already surpass other laboratory bounds. 
We further detail an experimental setup to directly probe this time-dependent signature via precision Mössbauer spectroscopy in the gravitational potential. This Letter demonstrates that this approach can significantly extend search capability and probe a vast, unexplored region of axion parameter space. Particularly, this setup can probe axion masses beyond the reach of existing experiments, such as atomic-clock measurements, offering a powerful new way for exploring higher-mass axion dark matter.
The sensitivity has the potential to be further improved with advancing experimental capabilities. 

\end{abstract}

\maketitle

\setcounter{secnumdepth}{3}
\setcounter{tocdepth}{1}

{\bf Introduction.}---%
The strong CP problem in Standard Model (SM) can be elegantly solved by introducing a Peccei-Quinn (PQ) symmetry, 
which is spontaneously broken at the $f_a$ scale. The associated Nambu-Goldstone boson, known as the QCD axion \cite{Peccei:1977hh,Peccei:1977ur,Weinberg:1977ma,Wilczek:1977pj,Shifman:1979if,Kim:1979if,Dine:1981rt,Zhitnitsky:1980tq}, is coupled to the gluon in terms of the dimension-5 operator. 
In the infrared (IR),  the nonperturbative dynamics from the QCD confinement induces the potential for the axion, generating a tiny mass given by $m_{a,{\rm QCD}} \simeq 6\mu\eV(10^{12}\GeV/f_a)$. 
We refer to axions satisfying this relation as the minimal realization of “QCD axions.” 
More generally, the QCD axion can solve the strong CP problem,  but with its mass deviating from the conventional mass relation.
In the following, we do not distinguish between the two and refer to both simply as axion in terms of the axion mass $m_a$ and decay constant $f_a$.

The axion can serve as a compelling dark matter (DM) candidate~\cite{Preskill:1982cy,Abbott:1982af,Dine:1982ah}. 
The tiny mass below eV scale implies that the axion behaves more wave-like instead of particle. 
One can use classical wave equation to describe the axion field over space. 
The local DM density is expected to be around $\rho_{\rm DM,local} = 0.4 \GeV\, \cm^{-3}$ \cite{Jungman_1996,Read_2014}. Therefore, within one coherence region, the local axion field can be described as
\begin{equation}
    a(t,\Vec{x}) \approx \dfrac{\sqrt{2\rho_{\rm DM,local}}}{m_a} \sin\brr{\omega_a t - \Vec{p} \cdot \Vec{x}}
\end{equation}
with the energy $\omega_a \approx m_a (1+ v^2/2)$ and the momentum $p = m_a v$. The value of $v$ is around 300 km/s. 
Consequently, the axion background can lead to an effective oscillating $\bar{\theta}_{\rm QCD}$ value. Such an effect leads to many experiments aiming to measure the time dependence in various observables, such as the neutron electric dipole moment (nEDM) and atomic EDMs~\cite{Stadnik:2013raa,Roberts:2014cga,Roberts:2014dda,Chang:2017ruk,Schulthess:2022pbp,Schulthess:2022pbp} 
, nuclear spin precession~\cite{Abel:2017rtm,JacksonKimball:2017elr}, etc. 

At energies well below the QCD confinement scale, the relevant physics is described by a chiral Lagrangian involving mesons and nucleons. Consequently, the axion's couplings to gluons and quarks in the high-energy regime must be matched to its effective couplings to hadrons at low energies. Specifically, an oscillating $\bar\theta_{\rm QCD}$ value induces temporal variations in the masses of mesons and baryons, as well as in their interaction couplings.

These variations in fundamental physical quantities can, in turn, modify nuclear binding energies. Such mechanisms have been previously studied in nuclear physics contexts, where a sizable effective $\bar{\theta}_{\rm QCD}$ angle is known to drastically impact nuclear processes~\cite{Broggini:2024udi,Alda:2024xxa,Zhang:2023lem}, both during Big Bang Nucleosynthesis (BBN)~\cite{Blum:2014vsa} and in stellar environments~\cite{Hook:2017psm,Zhang:2021mks,Balkin:2022qer,Gomez-Banon:2024oux,Kumamoto:2024wjd}. In this letter, we propose to use the M\"ossbauer spectroscopy to probe the oscillating $\bar\theta_{\rm QCD}$ value, which can be further interpreted in terms of the axion parameter space. 

Discovered in 1958~\cite{Mossbauer:1958wsu,Mossbauer2}, the Mössbauer effect takes advantage of the recoil-free emission and absorption of nuclear gamma rays in solid-state systems, enabling the observation of ultra-narrow lines in nuclear decay spectra. This high-precision technique provides a novel avenue for detecting small variations in nuclear binding energy, such as those induced by the oscillating field of axion dark matter.

A particularly notable Mössbauer transition is the 88 keV resonance line of the $^{109}$Ag isotope, which corresponds to the decay from the first excited state to the ground state and has a lifetime of 58 seconds~\cite{Adhikari_2018, LR1962691}. Measurements of this line achieved a sensitivity of 30 times its natural linewidth as early as 1979~\cite{wildner:jpa-00218534}, with subsequent studies refining this limit down to approximately 7 times the natural width~\cite{Taylor:1988,Hoy:1990,Alpatov:1996bd,Alpatov:2007LF,Bayukov2009,Alpatov2008}.

In this work, we adopt this $^{109}$Ag transition as a benchmark to demonstrate the potential of Mössbauer spectroscopy as a powerful tool in the search for axion dark matter.

{\bf Axion Effects}---%
We start from the simple case that the axion only couples to the gluon in the perturbative QCD regime. Other operators like the axion coupling to the quark chiral current could at most affect our results by $O(1)$. Below QCD confinement scale, Chiral Perturbation Theory (ChPT) should be used to describe the interactions among hadrons.  The existence of the QCD axion sets the time average value of $\bar \theta_{\rm QCD}$ to zero. However, the axion dark matter leads to a non-zero time dependent axion field value, which further induces a time dependent theta angle denoted as $\theta_{a}(x) = a(x)/f_a$.

Within the framework of Chiral Lagrangian, a non-zero $\theta_a$ term induces corrections to hadron masses and the couplings between mesons and nucleons. Since nuclear interactions are mediated by virtual meson exchange, the resulting binding energy is sensitive to these changes. Consequently, a dynamical effective $\bar
\theta_{\rm QCD}$ can alter nuclear binding energies.

For the heavy nuclei, the binding energy is primarily controlled by a spin singlet and isospin singlet central potential.
One can write a scalar and a vector contribution~\cite{Ubaldi:2008nf},
\begin{equation}
    H = \frac{1}{2}\alpha_S (\bar{N}N)(\bar{N}N) + \frac{1}{2} \alpha_V (\bar{N}\gamma^\mu N) (\bar{N}\gamma_\mu N) \ ,
\end{equation}
with the attractive force for the scalar interaction and the repulsive force for the vector interaction.
The scalar and vector components correspond to exchange of the $\sigma(600)$ meson and the $\omega(783)$ meson respectively. One can define the following ratio
\begin{equation}
   \eta_S(\theta_a)=\dfrac{\alpha_S(\theta_a)}{\alpha_S(\theta_a = 0)}, \quad   \eta_V(\theta_a)=\dfrac{\alpha_V(\theta_a)}{\alpha_V(\theta_a = 0)} \ .
\end{equation}
The potential energy is quite sensitive to the low energy physics, especially the threshold determined by the pion mass. Let us take the scalar potential as an example. The parametrization of $\eta_S$ can be written as by~\cite{Damour:2007uv}
\begin{equation} \label{eq:eta}
    \eta_S = -0.4\,\frac{m_\pi^2(\theta_a)}{m_{\pi}^2(\theta_a = 0)} + 1.4 \simeq 1 + 0.044 \, \theta_a^2 \ .
\end{equation}
We note the binding energy shift from axion background field is proportional to $\theta_a^2$.
Similar argument can be made for the vector component as well. In the following, we come to compute the nuclear binding energy explicitly term by term.

{\bf Nuclear Relativistic Density Functional Theory}---%

The energy scale of a typical nuclear gamma decay is on the order of 10 MeV. In contrast, the 88 keV transition in 
$^{109}$Ag is exceptionally small, arising from an accidental cancellation among several large contributions to the nuclear binding energy. This cancellation is highly advantageous for our search. An oscillating axion field perturbs the underlying interactions and hadron masses, affecting each of the large binding energy components in an uncorrelated manner. This perturbation disrupts the precise cancellation, effectively amplifying the relative shift in the resonance peak position.

In this section, we demonstrate that the 88 keV line indeed results from such an accidental cancellation. We further justify that the axion-induced variation in the binding energy should be referenced to the much larger fundamental scale of the nuclear forces, approximately 100 MeV, rather than to the small 88 keV transition energy. This effectively enhances the experimental sensitivity to the axion signal.

Now we come to explicitly calculate the binding energies for the ground and excited states of $^{109}$Ag, using nuclear relativistic density functional theory (RDFT).
The basis of nuclear RDFT is the Hohenberg-Kohn theorem~\cite{Hohenberg1964PhysRev.136.B864}, which demonstrates that for a many-particle system, the local one-body density constitutes a ``basic variable," i.e., all properties of the system can be written as unique functionals of the local density.
This makes the RDFT applicable for medium and heavy nuclear properties, for which direct solutions are very difficult, if not impossible.
In nuclear RDFT, such a density functional can be constructed starting from a Lorentz covariant Lagrangian density, and the local density that minimizes the nuclear binding energy is determined by iteratively solving the relativistic Hartree-Bogoliubov (RHB) equation; see more details about the Lagrangian density and the RHB equation in the Supplementary Material.
In the Lagrangian density, the scalar coupling $\alpha_S$ simulates the nuclear force from the $\sigma$ meson exchange; the vector coupling $\alpha_V$ accounts for that from the $\omega$ meson; the isovector-vector coupling $\alpha_{TV}$ simulates the one induced by the $\rho$ meson.
Note that the coupling simulating the $\pi$ exchange is not explicitly included in the present RDFT, because it carries negative parity and breaks parity on the Hartree level~\cite{Serot1997IJMPE6.515}.
Nevertheless, the effects of the $\pi$ exchange are implicitly taken into account in the RDFT as follows~\cite{Ring1996PPNP.37.193}.
First, the $\alpha_S$ coupling can be understood as an approximation to a two-pion resonance in a mesonic picture, and the $\alpha_{TV}$ coupling accounts for the isospin-dependent nuclear force associated with the $\pi$ exchange.
Second, additional $\pi$ exchange effects not covered by $\sigma$ and $\rho$ mesons are incorporated through suitable adjustments of the Lagrangian parameters.

To obtain the ground-state and excited-state energies of $^{109}$Ag, we use the well-known PC-PK1 relativistic density functional~\cite{Zhao2012PhysRevC.82.054319}, which has been successfully applied to nuclear mass~\cite{Yang2021PhysRevC.104.054312,Zhang2022ADNDT144.101488,Guo2024ADNDT158.101661}, novel rotations~\cite{Zhao2011PhysRevLett.107.122501,Zhao2017PLB773.1,Wang2024PLB848.138346}, fission~\cite{Ren2022PhysRevLett.128.172501,Zhang2024PhysRevC.109.024316,Li2023PhysRevC.107.014303}, and weak decay~\cite{Wang2024Sci.Bull.2017,Wang2024PLB855.138796} studies, demonstrating high predictive power.
The RHB equation is solved using a three-dimensional harmonic oscillator basis in Cartesian coordinates with $10$ major shells, and our calculations are free from additional adjustable parameters.
The calculated ground-state binding energy of $^{109}$Ag based on the proton $2p_{1/2}$ configuration is $928.814$ MeV, and that of the excited state based on the proton $1g_{9/2}$ configuration is $929.298$ MeV.
These values are in reasonable agreement with the experimental data $931.723$ MeV and $931.635$ MeV, respectively~\cite{Wang2021AME}.
The RDFT-based energy difference between the excited and ground states is $-0.483$ MeV, deviating slightly from the experimental value of $0.088$ MeV.

Now we have demonstrated that RDFT calculations successfully reproduce this exceptionally small energy splitting. Given the accuracy in capturing the 88 keV transition, which is very small compared to the typical nuclear decay scale, we conclude that the RDFT framework provides a reliable description of the underlying physics for both the ground and excited states in $^{109}$Ag.

Next, based on these RDFT results, we will explicitly justify that this small energy difference indeed results from a precise cancellation of large contributions originating from several distinct coupling channels.

{\bf Justification of the cancellation in $^{109}$Ag}---%
\begin{table}[htbp]
  \centering
  \caption{Contributions of $\alpha_S$, $\alpha_V$, and $\alpha_{TV}$ channels to the ground (G.s.) and excited (Ex.) states in $^{109}$Ag, along with the energy differences $\Delta E_i$ between these two states induced by each channel.}
  \begin{tabular}{cccc}
  \hline\hline
    Channels   & $E^{\mathrm{G.s.}}_i$ (MeV) & $E^{\mathrm{Ex.}}_i$ (MeV) & $\Delta E_i$ (MeV) \\
  \hline 
    $\alpha_S$ &  -17623.540                  &  -17510.543              &  +112.997    \\
    $\alpha_V$ &  +13065.872                  &  +12992.183              &  -73.689     \\
    $\alpha_{TV}$ & +19.427                   &  +19.154                 &  -0.273      \\
  \hline\hline 
  \end{tabular}\label{Tab:Diff-E}
\end{table}
The Lagrangian density in the RDFT includes various coupling channels.
Therefore, the energy difference between the ground and excited states can be attributed to the distinct contributions of each channel.
Since the axion modifies the coupling constants of these channels independently, the scale of the energy difference between the ground and excited states in each coupling channel determines how tightly the axion decay constant can be constrained.
In Tab.~\ref{Tab:Diff-E}, the contributions of selected channels, i.e., $\alpha_S$, $\alpha_V$, and $\alpha_{TV}$, to the ground and excited states, together with the energy difference $\Delta E_i$ induced by each channel, are shown.
The contributions from other channels can be seen in the Supplementary Material.

As shown in Tab.~\ref{Tab:Diff-E}, the ground and excited states in $^{109}$Ag arise from the balance between the large attractive potential from $\alpha_S$ and the strong repulsive potential from $\alpha_V$, both of which are significantly larger than that from $\alpha_{TV}$. 
Interestingly, although the calculated total energy difference $\Delta E$ between the excited state and the ground state is only about $0.5$ MeV, the individual contributions from the $\alpha_S$ and $\alpha_V$ channels are two orders of magnitude larger than the total. 
The $\Delta E$ contribution from the $\alpha_S$ channel is $+112.997$ MeV, whereas that from $\alpha_V$ channel is $-73.689$ MeV, counteracting the effect of $\alpha_S$.
Therefore, based on the microscopic RDFT calculations, it can be concluded that the small energy difference between the ground state and the excited state in $^{109}$Ag originates from the near-complete cancellation between the large positive contribution from $\alpha_S$ and the large negative contribution from $\alpha_V$, along with subtle competition among other coupling channels, as more clearly shown in the Supplementary Material.

For a given coupling channel (e.g., $\alpha_S$ or $\alpha_V$), the contributions to both the ground state and the first excited state originate from the same underlying nuclear physics, such as $\sigma$ or $\rho$ meson exchanges. Therefore, it is reasonable to assume that the axion-induced corrections to these states are correlated within the same channel. Consequently, the change in the binding energy contribution from a single channel $i$ can be expressed as $(1+\epsilon_i)E_i^{\rm G.s.} - (1+\epsilon_i)E_i^{\rm Ex.} = (1+\epsilon_i)\Delta E_i$.

In contrast, the correction factors $\epsilon_i$ are expected to be uncorrelated across different channels, as they arise from distinct physical mechanisms. We therefore anticipate that different $\epsilon_i$ will vary from one another at least by an $\mathcal{O}(1)$ factor.

More explicitly, let us consider the binding energy shift  from the scalar exchange channel as an example. Using the notation in Eq. (\ref{eq:eta}), the $\theta_a$ dependence in the binding energy difference can be written as
\begin{equation}
\begin{split}
    \Delta & E_S(\theta_a)-\Delta E_S(0) \\
    =& [E_S^{\rm Ex.}(\theta_a)-E_S^{\rm G.s.}(\theta_a)] - [E_S^{\rm Ex.}(0)-E_S^{\rm G.s.}(0)] \\
    \simeq& [\eta_S(\theta_a)-\eta_S(0)] [E_S^{\rm Ex.}(0)-E_S^{\rm G.s.}(0)] \\
    \simeq& 0.044\ \theta_a^2\times 113\text{MeV} \\
    =& 5.0\,\theta_a^2\text{MeV} .
\end{split}
\end{equation}
We now use this relation to estimate the sensitivity of our proposed axion search.

{\bf Constraints and projections to the axion parameter space}---%
The first excited state of $^{109}$Ag is long-lived, with a mean lifetime of approximately 58 s. Its decay produces an 88 keV emission line with an ultra-narrow intrinsic width of $\Gamma_0 = 1.16 \times 10^{-17}\ \text{eV}$.

This transition has been studied via Mössbauer spectroscopy at liquid helium temperatures. However, due to experimental limitations, the best resolution achieved to date corresponds to a measured linewidth of approximately 16 times the intrinsic value, $\Gamma_0$.

We now propose a detailed experimental strategy to search for axion dark matter using Mössbauer spectroscopy with this specific 88 keV emission line from $^{109}$Ag. Assuming the source is placed at the origin and the absorber is located at $\vec d$ with a distance of $d$. 
The binding energy change during the time for the photon propagating from the source to the absorber is given by
\begin{equation}
\begin{split}
    \delta E_{\rm bind}(d) & = \Delta E_{\rm bind}(t+d, \Vec{d}) - \Delta E_{\rm bind}(t,0) \\
    & \approx 0.005 \, \theta_a^2\srr{\sin^2\brr{\phi_s + m_a d} - \sin^2\phi_s} \GeV
\end{split}
\end{equation}
with $\phi_s$ as the oscillation phase. Depending on the magnitude of $m_a d$, this can be approximately simplied as 
\begin{equation}\label{eq:bind_E_diff}
\begin{split}
   \dfrac{\delta E_{\rm bind}(d)}{0.005 \, \theta_a^2 \, \GeV}  
   &\approx  \,  \sin(2\phi_s + m_a d)\sin(m_a d) \\
&\approx   \left\{
    \begin{aligned}
    & \sin (2\phi_s+m_a d)\times m_a d , \, & m_a d \ll 1 \\
    & 2^{-1/2}, \, & m_a d \gg 1
    \end{aligned}
\right. . 
\end{split}
\end{equation}
We note that the binding energy oscillates at the frequency $2m_a$. For $m_a d \gg 1$, we take the time-averaged value of the phase factor as $2^{-1/2}$. Note that the larger distance $d$ is, the suppression starts from a lower axion mass.

It is known that the photon would undergo the frequency shift when propagating within the gravitational potential. 
Therefore the resonance absorption position can only locate at around the same height as the source. 
Assuming the source is located at $Z=0$ along the vertical direction. Several absorbers
and detectors are placed at the height close to 0 and with equal distance $d$ to the source. 

\begin{figure}[t!]
    \centering
    \includegraphics[width=0.95\linewidth]{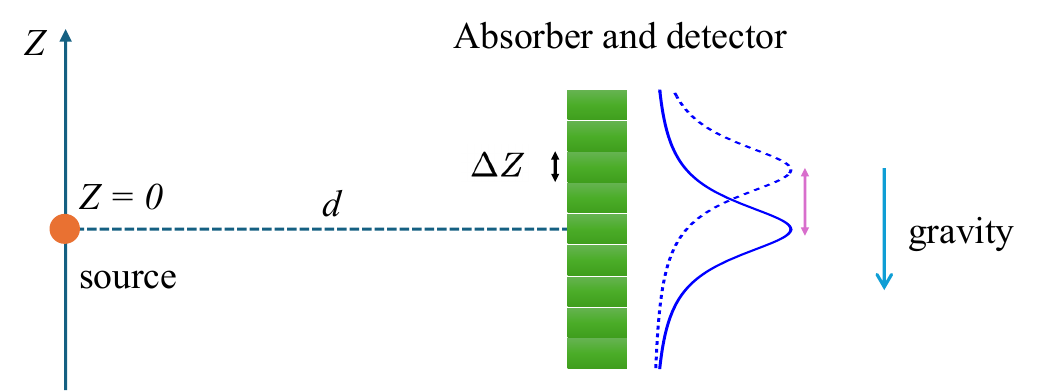}
    \caption{The illustrative experimental setup.}
    \label{fig:exp_illu}
\end{figure}

The illustrative experimental configuration is shown in Fig.~\ref{fig:exp_illu}. In addition, since the source emits photons isotropically, in order to improve the statistics, one can construct multiple copies of such absorber/detector configuration on the horizontal plane to form a  ring around the source.

The width of the absorbers is critically constrained. The gravitational redshift across the absorber's size must not cause a frequency broadening larger than the intrinsic linewidth of the Mössbauer transition. Quantitatively, the frequency resolution imposed by an absorber of width $\Delta Z$  is:
\begin{equation}
\frac{\delta f}{f} = g \, \Delta Z,
\end{equation}
where $g$ is the local gravitational acceleration.

To resolve the resonant absorption peak, the frequency resolution imposed by the detector width, $\Delta Z$, must be smaller than the intrinsic linewidth. For instance, in a terrestrial experiment aiming for a resolution of $\Gamma_0/2$, the detector width must not exceed 1$\mu$m.

As discussed above, the oscillating axion DM background can induce a time-dependent photon frequency shift.
In our experimental setup, this leads to a periodic modulation of the absorption peak position along the vertical direction, 
\begin{equation}
Z_p(t) = A \cos(\omega t + \phi).
\end{equation}
Our goal is to search for such an oscillatory feature in data and determine its frequency and amplitude. 
For a given amount of time $\delta T$, we find the number of events registered in each detector away from the resonance peak can be written as
\begin{equation}
    N_{\rm det} = R_s \, \text{Br} \, \delta T \dfrac{(2\pi d)\Delta Z}{4\pi d^2} \ ,
\end{equation}
where $R_s$ is the source radiation intensity in the unit of Ci. 
The factor Br refers to the branching ratio of the transition from the first excited state of $^{109}$Ag that produces the 88 keV photon, which has a value of 0.036~\cite{LEUTZ1965263,INL_Cd109}.
For the detectors close to the resonance peak, the number of registered photons have additional dependence on the recoil-free fraction of emitted photons from the source $f_S$,  as well as the resonant absorption efficiency of the absorber $\epsilon$. More details about how they affect the projection sensitivity can be found in the Supplementary Material.

The statistical uncertainty in each detector scales as $1/\sqrt{N_{\rm det}}$. As data in each detector accumulates, the statistical uncertainty will decrease, and systematic errors will eventually dominate the measurement. Potential sources of systematic error include lattice thermal vibrations at finite temperature, energy splitting from background magnetic fields, mechanical mismatch, source inhomogeneity, and detector instability etc.
Through careful calibration and tuning, these effects can be mitigated. For the following analysis, we will assume that the total systematic uncertainty is ultimately limited by the detector's frequency resolution, which is governed by the finite size of the absorber. In the Supplementary Material, we provide the explicit analysis strategy in order to balance the statistical and systematic uncertainties and to achieve the best sensitivity to the axion search.
\begin{table}[!t]
\begin{tabular}{|c|c|c|c|c|c|c|}
\hline
  & $g(g_\oplus)$ & d(m) & $\Delta Z$ & $\epsilon f_S$ & $m_{a, \rm max}$(eV) & $R_s$(Ci)      \\ \hline
A & 1     & 1    & 10$\mu$m  & 0.04 & $10^{-6}$ & $1$ \\ \hline
B & $10^{-4}$  & 100 & 1 dm & 0.04   & $10^{-6}$  & $10$ \\ \hline
\end{tabular}
\caption{The explicit parameters in two benchmark experimental configurations. Here $g_\oplus$ refers to the gravitational acceleration at the earth surface.}
\label{tab:exp}
\end{table}
\begin{figure*}[th]
    \centering
\includegraphics[width=0.7\textwidth]{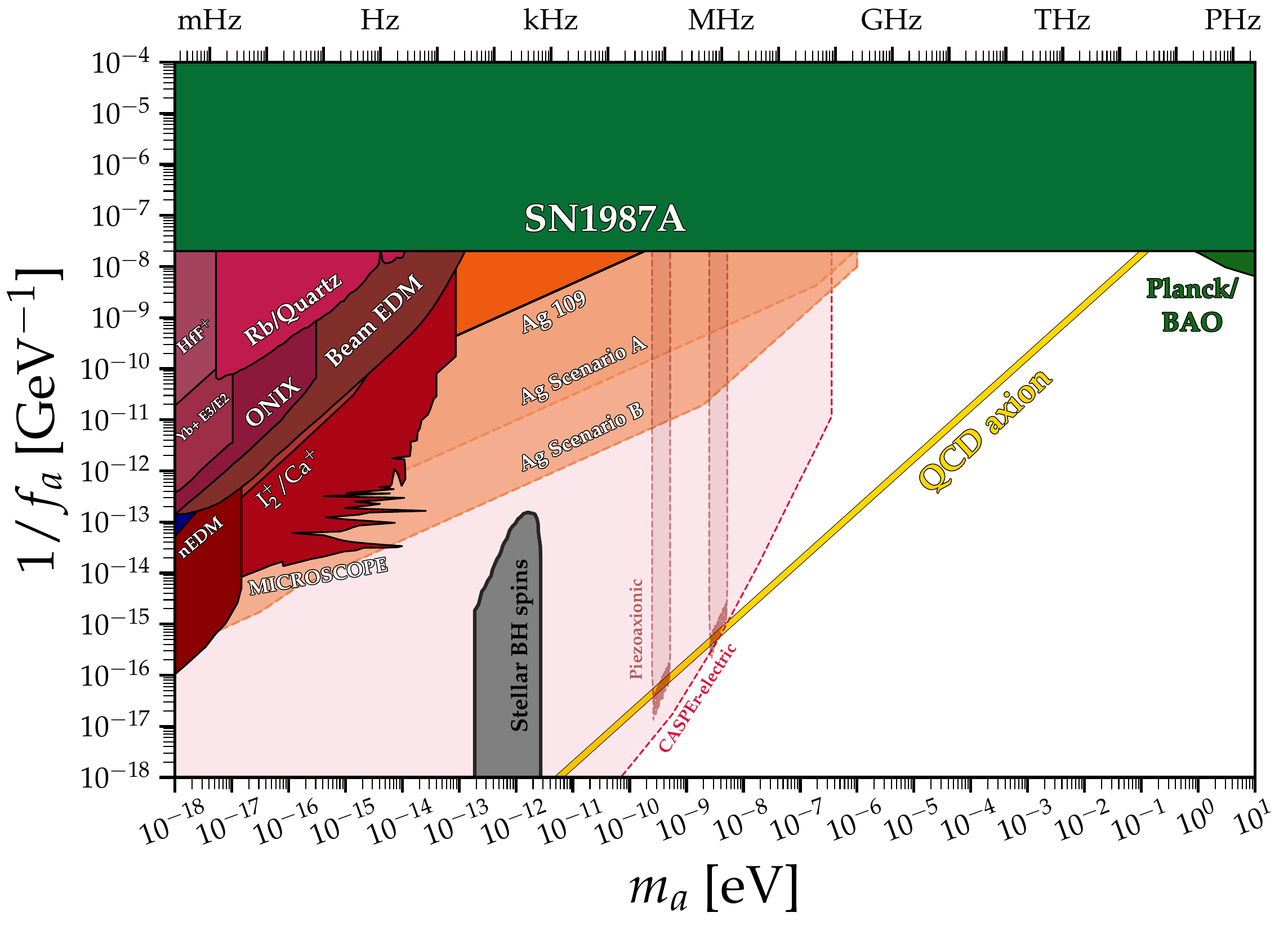}
    \caption{Here we show the sensitivity projection using $^{109}$Ag M\"ossbauer spectroscopy. The curve labeled ``Ag 109" is derived by reinterpreting existing measurements~\cite{Bayukov2009}. It is obtained by requiring that the axion-induced peak shift be smaller than the experimental linewidth precision of $7.0 \Gamma_0$, for a source-absorber separation of $d=
0.24$m. The two dashed lines represent the projected sensitivities in the benchmark Scenario A and B. Other experimental sensitivities on the    $f_a^{-1}-m_a$ plane are overlaid. }
    \label{fig:Constraint}
\end{figure*}

In this study, we work on two different benchmark  experimental configurations as in Tab.~\ref{tab:exp}. 
Scenario A is the setup of a terrestrial experiment. Scenario B is designed to be conducted in the near-earth outer space, following the suggestion in \cite{Gao:2023ggo} for the gravitational wave detection. It is reported that the current commercial technology can produce the samples with radioactive activity to be O(0.1-1) Ci. Hence we choose the source intensity to be 1 Ci in Scenario A  and 10 Ci in Scenario B as the benchmark values.
We choose the total observation duration to be 1 year. 
The projected sensitivity with 95\% C.L., in terms of the axion parameter space, is displayed in Fig.~\ref{fig:Constraint}.

In addition, we adopt a current estimation for a previous terrestrial experiment with a 24-centimeter baseline~\cite{Bayukov2009}. We require the axion-induced binding energy shift not to exceed the current experimental sensitivity, i.e. $\delta E_{\rm bind} \lesssim 7.0\, \Gamma_0$.

In the sensitivity projection, there are turning points corresponding to the critical point where $m_a d = 1$. 
The upper limit of the probed axion mass $m_a$ is determined by the detector time resolution $t_{\rm res}$, leading $m_a \leq 1\mu$eV in our benchmark value choice. Within this axion mass range that we can probe, the distance $d$ is always smaller than the axion coherence length.
The Supplementary Material provide a detailed description of the parameter space to which our analysis applies.

Comparing our projected sensitivities with other existing constraints~\cite{Schulthess:2022pbp,Abel:2017rtm,Roussy:2020ily,Madge:2024aot,Banerjee:2023bjc,Mehta:2020kwu,Baryakhtar:2020gao,Unal:2020jiy,Hoof:2024quk,Witte:2024drg} as well as the projected sensitivity of the CASPERr-electric~\cite{JacksonKimball:2017elr} and Piezoaxionic experiments~\cite{Arvanitaki:2021wjk}, our proposed search has the capability to probe interesting unexplored parameter space, even with the conservative experimental setup. 

It is instructive to compare our projected sensitivity with results from atomic clock experiments, such as the ${\rm I_2^+/Ca^+}$ system ~\cite{Madge:2024aot}. Both approaches use ultra-narrow spectral lines. The ${\rm Ca^+}$ clock transition has an intrinsic linewidth of $\Gamma_0(\rm Ca^+) = (1.16\ \text{s})^{-1} \approx 4.1 \times 10^{-16}$ eV, which is comparable to the $^{109}$Ag M\"ossbauer linewidth of $\Gamma_0 = 1.16 \times 10^{-17}$ eV. The critical difference lies in the transition energy. Atomic clock transitions occur at eV-scale energies, while our nuclear transition energy is 88 keV. This results in a fractional frequency sensitivity, $\delta f / f = \Gamma_0 / E_\gamma$, that is at least five orders of magnitude more favorable for the M\"ossbauer technique.

A second crucial difference lies in the intrinsic time resolution of the two methods. In an atomic clock, the time resolution is set by the interrogation time of each measurement.
This interval represents a fundamental trade-off. If it exceeds the excited state's lifetime, the accumulated signal suffers from decoherence; if it is too short, the photon-frequency uncertainty increases due to Fourier broadening. To optimize sensitivity, the interrogation time is typically chosen to be approximately half the excited-state lifetime, which corresponds to roughly O(0.1) seconds in standard configurations. In contrast, Mössbauer spectroscopy relies on photon counting, where the upper frequency limit is primarily set by the detector's time resolution, typically on the nanosecond scale. This corresponds to a sampling frequency approximately eight orders of magnitude higher. This higher sampling frequency enables the probe of axion masses significantly larger than those accessible to atomic clock experiments.

{\bf Possible UV Completion}---%
Similar to other QCD axion searches, the parameter space currently accessible to our proposed design requires an axion mass lower than that predicted by minimal QCD axion models, for a fixed gluon coupling. However, there are few UV-complete models that naturally explain such a suppressed axion mass while maintaining the coupling. One example is the $\mathbb{Z}_N$ model~\cite{Hook:2018jle,DiLuzio:2021pxd}, which reduces the axion mass via contributions from multiple dark sectors. Nevertheless, this model faces severe constraints from finite-density effects. In dense macroscopic objects, medium effects can induce an additional axion potential that flips the $\bar{\theta}_{\rm QCD}$ vacuum from 0 to $\pi$, and a large portion of the model's parameter space is thus excluded by measurements of the Sun, white dwarfs, and neutron stars.

Conversely, other UV-complete models can readily evade these astrophysical constraints while still predicting a suppressed axion mass. For instance, the model in~\cite{Co:2024bme} introduces a contribution to the axion potential from a dark sector, leading to a significant cancellation in the second derivative of the potential at the minimum ($\bar \theta_{\rm QCD}=0$). This cancellation results in a low mass at the vacuum.

However, this cancellation is specific to the potential minimum. The overall scale of the potential away from $\bar \theta_{\rm QCD}=0$ remains comparable to that of the conventional QCD axion. Consequently, within dense stellar environments, finite-density effects would only shift the $\bar \theta_{\rm QCD}$ slightly from its origin. This minimal displacement is expected to produce negligible modifications to stellar properties, thereby avoiding the stringent constraints faced by other models.

In this study, we do not specify the explicit models to suppress the QCD axion mass. 
Therefore we drop the constraints from the Sun, white dwarfs and neutron stars.

{\bf Conclusion and Outlook}---%

In this letter, we propose to search for the axion dark matter using M\"ossbauer spectroscopy via the 88 keV  emission line from the first excited state of  $^{109}$Ag. 
The ultra-narrow linewidth provides a highly sensitive probe of the binding energy shift induced by the oscillating axion background field. Existing high-precision measurements of the photon linewidth from previous M\"ossbauer spectroscopy experiments can be directly reinterpreted to constrain the QCD axion's mass and decay constant. Furthermore, we demonstrate that the axion-induced binding energy shift manifests as a displacement of the resonance peak along the vertical direction due to the gravitational potential. By vertically separating the source and absorber/detector, a M\"ossbauer spectroscopy experiment can be used to probe axion dark matter.
Here we present our analysis of the projected sensitivity based on several benchmark experimental configurations.
Our results indicate that  a broad range of axion parameters can be safely covered. 
M\"ossbauer spectroscopy offers a unique advantage in probing higher axion masses compared to atomic experiments, due to its inherently high time resolution.
Further improvements, such as detector resolution and the source radioactivity, could significantly enhance the  precision and extend the search  for axion dark matter.

{\bf Acknowledgment}\label{sec:acknowledgement}
We thank Huaqiao Zhang for useful discussions. K-F.L would like to thank the Aspen Center for Physics, which is supported by National
Science Foundation grant PHY-2210452, for hospitality during the course of this work. J.M. is supported by the National Natural Science Foundation of China (NSFC) under Grants No. 12435006, the National Key Laboratory of Neutron Science and Technology (Grants No. NST202401016), the National Key R$\&$D Program of China (Grant No. 2024YFE0109803), and the State Key Laboratory of Nuclear Physics and Technology, Peking University (Grant No. NPT2023ZX03).
J.S. is supported by Peking University under startup Grant No. 7101302974 and the NSFC under Grants No. 12025507, No.12450006.
Y.W. is supported by the NSFC under Grants No. 12141501, No. 12105004, the Beijing Natural Science Foundation (Grant No. 1242019).
K-F.L and Y.Z. are supported by U.S. Department of Energy under Award No. DESC0009959.

\bibliography{ref}

\begin{thebibliography}{74}%
\makeatletter
\providecommand \@ifxundefined [1]{%
 \@ifx{#1\undefined}
}%
\providecommand \@ifnum [1]{%
 \ifnum #1\expandafter \@firstoftwo
 \else \expandafter \@secondoftwo
 \fi
}%
\providecommand \@ifx [1]{%
 \ifx #1\expandafter \@firstoftwo
 \else \expandafter \@secondoftwo
 \fi
}%
\providecommand \natexlab [1]{#1}%
\providecommand \enquote  [1]{``#1''}%
\providecommand \bibnamefont  [1]{#1}%
\providecommand \bibfnamefont [1]{#1}%
\providecommand \citenamefont [1]{#1}%
\providecommand \href@noop [0]{\@secondoftwo}%
\providecommand \href [0]{\begingroup \@sanitize@url \@href}%
\providecommand \@href[1]{\@@startlink{#1}\@@href}%
\providecommand \@@href[1]{\endgroup#1\@@endlink}%
\providecommand \@sanitize@url [0]{\catcode `\\12\catcode `\$12\catcode `\&12\catcode `\#12\catcode `\^12\catcode `\_12\catcode `\%12\relax}%
\providecommand \@@startlink[1]{}%
\providecommand \@@endlink[0]{}%
\providecommand \url  [0]{\begingroup\@sanitize@url \@url }%
\providecommand \@url [1]{\endgroup\@href {#1}{\urlprefix }}%
\providecommand \urlprefix  [0]{URL }%
\providecommand \Eprint [0]{\href }%
\providecommand \doibase [0]{http://dx.doi.org/}%
\providecommand \selectlanguage [0]{\@gobble}%
\providecommand \bibinfo  [0]{\@secondoftwo}%
\providecommand \bibfield  [0]{\@secondoftwo}%
\providecommand \translation [1]{[#1]}%
\providecommand \BibitemOpen [0]{}%
\providecommand \bibitemStop [0]{}%
\providecommand \bibitemNoStop [0]{.\EOS\space}%
\providecommand \EOS [0]{\spacefactor3000\relax}%
\providecommand \BibitemShut  [1]{\csname bibitem#1\endcsname}%
\let\auto@bib@innerbib\@empty
\bibitem [{\citenamefont {Peccei}\ and\ \citenamefont {Quinn}(1977{\natexlab{a}})}]{Peccei:1977hh}%
  \BibitemOpen
  \bibfield  {author} {\bibinfo {author} {\bibfnamefont {R.~D.}\ \bibnamefont {Peccei}}\ and\ \bibinfo {author} {\bibfnamefont {H.~R.}\ \bibnamefont {Quinn}},\ }\href {\doibase 10.1103/PhysRevLett.38.1440} {\bibfield  {journal} {\bibinfo  {journal} {Phys. Rev. Lett.}\ }\textbf {\bibinfo {volume} {38}},\ \bibinfo {pages} {1440} (\bibinfo {year} {1977}{\natexlab{a}})}\BibitemShut {NoStop}%
\bibitem [{\citenamefont {Peccei}\ and\ \citenamefont {Quinn}(1977{\natexlab{b}})}]{Peccei:1977ur}%
  \BibitemOpen
  \bibfield  {author} {\bibinfo {author} {\bibfnamefont {R.~D.}\ \bibnamefont {Peccei}}\ and\ \bibinfo {author} {\bibfnamefont {H.~R.}\ \bibnamefont {Quinn}},\ }\href {\doibase 10.1103/PhysRevD.16.1791} {\bibfield  {journal} {\bibinfo  {journal} {Phys. Rev. D}\ }\textbf {\bibinfo {volume} {16}},\ \bibinfo {pages} {1791} (\bibinfo {year} {1977}{\natexlab{b}})}\BibitemShut {NoStop}%
\bibitem [{\citenamefont {Weinberg}(1978)}]{Weinberg:1977ma}%
  \BibitemOpen
  \bibfield  {author} {\bibinfo {author} {\bibfnamefont {S.}~\bibnamefont {Weinberg}},\ }\href {\doibase 10.1103/PhysRevLett.40.223} {\bibfield  {journal} {\bibinfo  {journal} {Phys. Rev. Lett.}\ }\textbf {\bibinfo {volume} {40}},\ \bibinfo {pages} {223} (\bibinfo {year} {1978})}\BibitemShut {NoStop}%
\bibitem [{\citenamefont {Wilczek}(1978)}]{Wilczek:1977pj}%
  \BibitemOpen
  \bibfield  {author} {\bibinfo {author} {\bibfnamefont {F.}~\bibnamefont {Wilczek}},\ }\href {\doibase 10.1103/PhysRevLett.40.279} {\bibfield  {journal} {\bibinfo  {journal} {Phys. Rev. Lett.}\ }\textbf {\bibinfo {volume} {40}},\ \bibinfo {pages} {279} (\bibinfo {year} {1978})}\BibitemShut {NoStop}%
\bibitem [{\citenamefont {Shifman}\ \emph {et~al.}(1980)\citenamefont {Shifman}, \citenamefont {Vainshtein},\ and\ \citenamefont {Zakharov}}]{Shifman:1979if}%
  \BibitemOpen
  \bibfield  {author} {\bibinfo {author} {\bibfnamefont {M.~A.}\ \bibnamefont {Shifman}}, \bibinfo {author} {\bibfnamefont {A.~I.}\ \bibnamefont {Vainshtein}}, \ and\ \bibinfo {author} {\bibfnamefont {V.~I.}\ \bibnamefont {Zakharov}},\ }\href {\doibase 10.1016/0550-3213(80)90209-6} {\bibfield  {journal} {\bibinfo  {journal} {Nucl. Phys. B}\ }\textbf {\bibinfo {volume} {166}},\ \bibinfo {pages} {493} (\bibinfo {year} {1980})}\BibitemShut {NoStop}%
\bibitem [{\citenamefont {Kim}(1979)}]{Kim:1979if}%
  \BibitemOpen
  \bibfield  {author} {\bibinfo {author} {\bibfnamefont {J.~E.}\ \bibnamefont {Kim}},\ }\href {\doibase 10.1103/PhysRevLett.43.103} {\bibfield  {journal} {\bibinfo  {journal} {Phys. Rev. Lett.}\ }\textbf {\bibinfo {volume} {43}},\ \bibinfo {pages} {103} (\bibinfo {year} {1979})}\BibitemShut {NoStop}%
\bibitem [{\citenamefont {Dine}\ \emph {et~al.}(1981)\citenamefont {Dine}, \citenamefont {Fischler},\ and\ \citenamefont {Srednicki}}]{Dine:1981rt}%
  \BibitemOpen
  \bibfield  {author} {\bibinfo {author} {\bibfnamefont {M.}~\bibnamefont {Dine}}, \bibinfo {author} {\bibfnamefont {W.}~\bibnamefont {Fischler}}, \ and\ \bibinfo {author} {\bibfnamefont {M.}~\bibnamefont {Srednicki}},\ }\href {\doibase 10.1016/0370-2693(81)90590-6} {\bibfield  {journal} {\bibinfo  {journal} {Phys. Lett. B}\ }\textbf {\bibinfo {volume} {104}},\ \bibinfo {pages} {199} (\bibinfo {year} {1981})}\BibitemShut {NoStop}%
\bibitem [{\citenamefont {Zhitnitsky}(1980)}]{Zhitnitsky:1980tq}%
  \BibitemOpen
  \bibfield  {author} {\bibinfo {author} {\bibfnamefont {A.~R.}\ \bibnamefont {Zhitnitsky}},\ }\href@noop {} {\bibfield  {journal} {\bibinfo  {journal} {Sov. J. Nucl. Phys.}\ }\textbf {\bibinfo {volume} {31}},\ \bibinfo {pages} {260} (\bibinfo {year} {1980})}\BibitemShut {NoStop}%
\bibitem [{\citenamefont {Preskill}\ \emph {et~al.}(1983)\citenamefont {Preskill}, \citenamefont {Wise},\ and\ \citenamefont {Wilczek}}]{Preskill:1982cy}%
  \BibitemOpen
  \bibfield  {author} {\bibinfo {author} {\bibfnamefont {J.}~\bibnamefont {Preskill}}, \bibinfo {author} {\bibfnamefont {M.~B.}\ \bibnamefont {Wise}}, \ and\ \bibinfo {author} {\bibfnamefont {F.}~\bibnamefont {Wilczek}},\ }\href {\doibase 10.1016/0370-2693(83)90637-8} {\bibfield  {journal} {\bibinfo  {journal} {Phys. Lett. B}\ }\textbf {\bibinfo {volume} {120}},\ \bibinfo {pages} {127} (\bibinfo {year} {1983})}\BibitemShut {NoStop}%
\bibitem [{\citenamefont {Abbott}\ and\ \citenamefont {Sikivie}(1983)}]{Abbott:1982af}%
  \BibitemOpen
  \bibfield  {author} {\bibinfo {author} {\bibfnamefont {L.~F.}\ \bibnamefont {Abbott}}\ and\ \bibinfo {author} {\bibfnamefont {P.}~\bibnamefont {Sikivie}},\ }\href {\doibase 10.1016/0370-2693(83)90638-X} {\bibfield  {journal} {\bibinfo  {journal} {Phys. Lett. B}\ }\textbf {\bibinfo {volume} {120}},\ \bibinfo {pages} {133} (\bibinfo {year} {1983})}\BibitemShut {NoStop}%
\bibitem [{\citenamefont {Dine}\ and\ \citenamefont {Fischler}(1983)}]{Dine:1982ah}%
  \BibitemOpen
  \bibfield  {author} {\bibinfo {author} {\bibfnamefont {M.}~\bibnamefont {Dine}}\ and\ \bibinfo {author} {\bibfnamefont {W.}~\bibnamefont {Fischler}},\ }\href {\doibase 10.1016/0370-2693(83)90639-1} {\bibfield  {journal} {\bibinfo  {journal} {Phys. Lett. B}\ }\textbf {\bibinfo {volume} {120}},\ \bibinfo {pages} {137} (\bibinfo {year} {1983})}\BibitemShut {NoStop}%
\bibitem [{\citenamefont {Jungman}\ \emph {et~al.}(1996)\citenamefont {Jungman}, \citenamefont {Kamionkowski},\ and\ \citenamefont {Griest}}]{Jungman_1996}%
  \BibitemOpen
  \bibfield  {author} {\bibinfo {author} {\bibfnamefont {G.}~\bibnamefont {Jungman}}, \bibinfo {author} {\bibfnamefont {M.}~\bibnamefont {Kamionkowski}}, \ and\ \bibinfo {author} {\bibfnamefont {K.}~\bibnamefont {Griest}},\ }\href {\doibase 10.1016/0370-1573(95)00058-5} {\bibfield  {journal} {\bibinfo  {journal} {Physics Reports}\ }\textbf {\bibinfo {volume} {267}},\ \bibinfo {pages} {195–373} (\bibinfo {year} {1996})}\BibitemShut {NoStop}%
\bibitem [{\citenamefont {Read}(2014)}]{Read_2014}%
  \BibitemOpen
  \bibfield  {author} {\bibinfo {author} {\bibfnamefont {J.~I.}\ \bibnamefont {Read}},\ }\href {\doibase 10.1088/0954-3899/41/6/063101} {\bibfield  {journal} {\bibinfo  {journal} {Journal of Physics G: Nuclear and Particle Physics}\ }\textbf {\bibinfo {volume} {41}},\ \bibinfo {pages} {063101} (\bibinfo {year} {2014})}\BibitemShut {NoStop}%
\bibitem [{\citenamefont {Stadnik}\ and\ \citenamefont {Flambaum}(2014)}]{Stadnik:2013raa}%
  \BibitemOpen
  \bibfield  {author} {\bibinfo {author} {\bibfnamefont {Y.~V.}\ \bibnamefont {Stadnik}}\ and\ \bibinfo {author} {\bibfnamefont {V.~V.}\ \bibnamefont {Flambaum}},\ }\href {\doibase 10.1103/PhysRevD.89.043522} {\bibfield  {journal} {\bibinfo  {journal} {Phys. Rev. D}\ }\textbf {\bibinfo {volume} {89}},\ \bibinfo {pages} {043522} (\bibinfo {year} {2014})},\ \Eprint {http://arxiv.org/abs/1312.6667} {arXiv:1312.6667 [hep-ph]} \BibitemShut {NoStop}%
\bibitem [{\citenamefont {Roberts}\ \emph {et~al.}(2014{\natexlab{a}})\citenamefont {Roberts}, \citenamefont {Stadnik}, \citenamefont {Dzuba}, \citenamefont {Flambaum}, \citenamefont {Leefer},\ and\ \citenamefont {Budker}}]{Roberts:2014cga}%
  \BibitemOpen
  \bibfield  {author} {\bibinfo {author} {\bibfnamefont {B.~M.}\ \bibnamefont {Roberts}}, \bibinfo {author} {\bibfnamefont {Y.~V.}\ \bibnamefont {Stadnik}}, \bibinfo {author} {\bibfnamefont {V.~A.}\ \bibnamefont {Dzuba}}, \bibinfo {author} {\bibfnamefont {V.~V.}\ \bibnamefont {Flambaum}}, \bibinfo {author} {\bibfnamefont {N.}~\bibnamefont {Leefer}}, \ and\ \bibinfo {author} {\bibfnamefont {D.}~\bibnamefont {Budker}},\ }\href {\doibase 10.1103/PhysRevD.90.096005} {\bibfield  {journal} {\bibinfo  {journal} {Phys. Rev. D}\ }\textbf {\bibinfo {volume} {90}},\ \bibinfo {pages} {096005} (\bibinfo {year} {2014}{\natexlab{a}})},\ \Eprint {http://arxiv.org/abs/1409.2564} {arXiv:1409.2564 [hep-ph]} \BibitemShut {NoStop}%
\bibitem [{\citenamefont {Roberts}\ \emph {et~al.}(2014{\natexlab{b}})\citenamefont {Roberts}, \citenamefont {Stadnik}, \citenamefont {Dzuba}, \citenamefont {Flambaum}, \citenamefont {Leefer},\ and\ \citenamefont {Budker}}]{Roberts:2014dda}%
  \BibitemOpen
  \bibfield  {author} {\bibinfo {author} {\bibfnamefont {B.~M.}\ \bibnamefont {Roberts}}, \bibinfo {author} {\bibfnamefont {Y.~V.}\ \bibnamefont {Stadnik}}, \bibinfo {author} {\bibfnamefont {V.~A.}\ \bibnamefont {Dzuba}}, \bibinfo {author} {\bibfnamefont {V.~V.}\ \bibnamefont {Flambaum}}, \bibinfo {author} {\bibfnamefont {N.}~\bibnamefont {Leefer}}, \ and\ \bibinfo {author} {\bibfnamefont {D.}~\bibnamefont {Budker}},\ }\href {\doibase 10.1103/PhysRevLett.113.081601} {\bibfield  {journal} {\bibinfo  {journal} {Phys. Rev. Lett.}\ }\textbf {\bibinfo {volume} {113}},\ \bibinfo {pages} {081601} (\bibinfo {year} {2014}{\natexlab{b}})},\ \Eprint {http://arxiv.org/abs/1404.2723} {arXiv:1404.2723 [hep-ph]} \BibitemShut {NoStop}%
\bibitem [{\citenamefont {Chang}\ \emph {et~al.}(2019)\citenamefont {Chang}, \citenamefont {Haciomeroglu}, \citenamefont {Kim}, \citenamefont {Lee}, \citenamefont {Park},\ and\ \citenamefont {Semertzidis}}]{Chang:2017ruk}%
  \BibitemOpen
  \bibfield  {author} {\bibinfo {author} {\bibfnamefont {S.~P.}\ \bibnamefont {Chang}}, \bibinfo {author} {\bibfnamefont {S.}~\bibnamefont {Haciomeroglu}}, \bibinfo {author} {\bibfnamefont {O.}~\bibnamefont {Kim}}, \bibinfo {author} {\bibfnamefont {S.}~\bibnamefont {Lee}}, \bibinfo {author} {\bibfnamefont {S.}~\bibnamefont {Park}}, \ and\ \bibinfo {author} {\bibfnamefont {Y.~K.}\ \bibnamefont {Semertzidis}},\ }\href {\doibase 10.1103/PhysRevD.99.083002} {\bibfield  {journal} {\bibinfo  {journal} {Phys. Rev. D}\ }\textbf {\bibinfo {volume} {99}},\ \bibinfo {pages} {083002} (\bibinfo {year} {2019})},\ \Eprint {http://arxiv.org/abs/1710.05271} {arXiv:1710.05271 [hep-ex]} \BibitemShut {NoStop}%
\bibitem [{\citenamefont {Schulthess}\ \emph {et~al.}(2022)\citenamefont {Schulthess} \emph {et~al.}}]{Schulthess:2022pbp}%
  \BibitemOpen
  \bibfield  {author} {\bibinfo {author} {\bibfnamefont {I.}~\bibnamefont {Schulthess}} \emph {et~al.},\ }\href {\doibase 10.1103/PhysRevLett.129.191801} {\bibfield  {journal} {\bibinfo  {journal} {Phys. Rev. Lett.}\ }\textbf {\bibinfo {volume} {129}},\ \bibinfo {pages} {191801} (\bibinfo {year} {2022})},\ \Eprint {http://arxiv.org/abs/2204.01454} {arXiv:2204.01454 [hep-ex]} \BibitemShut {NoStop}%
\bibitem [{\citenamefont {Abel}\ \emph {et~al.}(2017)\citenamefont {Abel} \emph {et~al.}}]{Abel:2017rtm}%
  \BibitemOpen
  \bibfield  {author} {\bibinfo {author} {\bibfnamefont {C.}~\bibnamefont {Abel}} \emph {et~al.},\ }\href {\doibase 10.1103/PhysRevX.7.041034} {\bibfield  {journal} {\bibinfo  {journal} {Phys. Rev. X}\ }\textbf {\bibinfo {volume} {7}},\ \bibinfo {pages} {041034} (\bibinfo {year} {2017})},\ \Eprint {http://arxiv.org/abs/1708.06367} {arXiv:1708.06367 [hep-ph]} \BibitemShut {NoStop}%
\bibitem [{\citenamefont {Jackson~Kimball}\ \emph {et~al.}(2020)\citenamefont {Jackson~Kimball} \emph {et~al.}}]{JacksonKimball:2017elr}%
  \BibitemOpen
  \bibfield  {author} {\bibinfo {author} {\bibfnamefont {D.~F.}\ \bibnamefont {Jackson~Kimball}} \emph {et~al.},\ }\href {\doibase 10.1007/978-3-030-43761-9_13} {\bibfield  {journal} {\bibinfo  {journal} {Springer Proc. Phys.}\ }\textbf {\bibinfo {volume} {245}},\ \bibinfo {pages} {105} (\bibinfo {year} {2020})},\ \Eprint {http://arxiv.org/abs/1711.08999} {arXiv:1711.08999 [physics.ins-det]} \BibitemShut {NoStop}%
\bibitem [{\citenamefont {Broggini}\ \emph {et~al.}(2024)\citenamefont {Broggini}, \citenamefont {Di~Carlo}, \citenamefont {Di~Luzio},\ and\ \citenamefont {Toni}}]{Broggini:2024udi}%
  \BibitemOpen
  \bibfield  {author} {\bibinfo {author} {\bibfnamefont {C.}~\bibnamefont {Broggini}}, \bibinfo {author} {\bibfnamefont {G.}~\bibnamefont {Di~Carlo}}, \bibinfo {author} {\bibfnamefont {L.}~\bibnamefont {Di~Luzio}}, \ and\ \bibinfo {author} {\bibfnamefont {C.}~\bibnamefont {Toni}},\ }\href {\doibase 10.1016/j.physletb.2024.138836} {\bibfield  {journal} {\bibinfo  {journal} {Phys. Lett. B}\ }\textbf {\bibinfo {volume} {855}},\ \bibinfo {pages} {138836} (\bibinfo {year} {2024})},\ \Eprint {http://arxiv.org/abs/2404.18993} {arXiv:2404.18993 [hep-ph]} \BibitemShut {NoStop}%
\bibitem [{\citenamefont {Alda}\ \emph {et~al.}(2025)\citenamefont {Alda}, \citenamefont {Broggini}, \citenamefont {Di~Carlo}, \citenamefont {Di~Luzio}, \citenamefont {Piatti}, \citenamefont {Rigolin},\ and\ \citenamefont {Toni}}]{Alda:2024xxa}%
  \BibitemOpen
  \bibfield  {author} {\bibinfo {author} {\bibfnamefont {J.}~\bibnamefont {Alda}}, \bibinfo {author} {\bibfnamefont {C.}~\bibnamefont {Broggini}}, \bibinfo {author} {\bibfnamefont {G.}~\bibnamefont {Di~Carlo}}, \bibinfo {author} {\bibfnamefont {L.}~\bibnamefont {Di~Luzio}}, \bibinfo {author} {\bibfnamefont {D.}~\bibnamefont {Piatti}}, \bibinfo {author} {\bibfnamefont {S.}~\bibnamefont {Rigolin}}, \ and\ \bibinfo {author} {\bibfnamefont {C.}~\bibnamefont {Toni}},\ }\href {\doibase 10.1103/PhysRevD.111.035022} {\bibfield  {journal} {\bibinfo  {journal} {Phys. Rev. D}\ }\textbf {\bibinfo {volume} {111}},\ \bibinfo {pages} {035022} (\bibinfo {year} {2025})},\ \Eprint {http://arxiv.org/abs/2412.20932} {arXiv:2412.20932 [hep-ph]} \BibitemShut {NoStop}%
\bibitem [{\citenamefont {Zhang}\ \emph {et~al.}(2023)\citenamefont {Zhang}, \citenamefont {Houston},\ and\ \citenamefont {Li}}]{Zhang:2023lem}%
  \BibitemOpen
  \bibfield  {author} {\bibinfo {author} {\bibfnamefont {X.}~\bibnamefont {Zhang}}, \bibinfo {author} {\bibfnamefont {N.}~\bibnamefont {Houston}}, \ and\ \bibinfo {author} {\bibfnamefont {T.}~\bibnamefont {Li}},\ }\href {\doibase 10.1103/PhysRevD.108.L071101} {\bibfield  {journal} {\bibinfo  {journal} {Phys. Rev. D}\ }\textbf {\bibinfo {volume} {108}},\ \bibinfo {pages} {L071101} (\bibinfo {year} {2023})},\ \Eprint {http://arxiv.org/abs/2303.09865} {arXiv:2303.09865 [hep-ph]} \BibitemShut {NoStop}%
\bibitem [{\citenamefont {Blum}\ \emph {et~al.}(2014)\citenamefont {Blum}, \citenamefont {D'Agnolo}, \citenamefont {Lisanti},\ and\ \citenamefont {Safdi}}]{Blum:2014vsa}%
  \BibitemOpen
  \bibfield  {author} {\bibinfo {author} {\bibfnamefont {K.}~\bibnamefont {Blum}}, \bibinfo {author} {\bibfnamefont {R.~T.}\ \bibnamefont {D'Agnolo}}, \bibinfo {author} {\bibfnamefont {M.}~\bibnamefont {Lisanti}}, \ and\ \bibinfo {author} {\bibfnamefont {B.~R.}\ \bibnamefont {Safdi}},\ }\href {\doibase 10.1016/j.physletb.2014.07.059} {\bibfield  {journal} {\bibinfo  {journal} {Phys. Lett. B}\ }\textbf {\bibinfo {volume} {737}},\ \bibinfo {pages} {30} (\bibinfo {year} {2014})},\ \Eprint {http://arxiv.org/abs/1401.6460} {arXiv:1401.6460 [hep-ph]} \BibitemShut {NoStop}%
\bibitem [{\citenamefont {Hook}\ and\ \citenamefont {Huang}(2018)}]{Hook:2017psm}%
  \BibitemOpen
  \bibfield  {author} {\bibinfo {author} {\bibfnamefont {A.}~\bibnamefont {Hook}}\ and\ \bibinfo {author} {\bibfnamefont {J.}~\bibnamefont {Huang}},\ }\href {\doibase 10.1007/JHEP06(2018)036} {\bibfield  {journal} {\bibinfo  {journal} {JHEP}\ }\textbf {\bibinfo {volume} {06}},\ \bibinfo {pages} {036} (\bibinfo {year} {2018})},\ \Eprint {http://arxiv.org/abs/1708.08464} {arXiv:1708.08464 [hep-ph]} \BibitemShut {NoStop}%
\bibitem [{\citenamefont {Zhang}\ \emph {et~al.}(2021)\citenamefont {Zhang}, \citenamefont {Lyu}, \citenamefont {Huang}, \citenamefont {Johnson}, \citenamefont {Sagunski}, \citenamefont {Sakellariadou},\ and\ \citenamefont {Yang}}]{Zhang:2021mks}%
  \BibitemOpen
  \bibfield  {author} {\bibinfo {author} {\bibfnamefont {J.}~\bibnamefont {Zhang}}, \bibinfo {author} {\bibfnamefont {Z.}~\bibnamefont {Lyu}}, \bibinfo {author} {\bibfnamefont {J.}~\bibnamefont {Huang}}, \bibinfo {author} {\bibfnamefont {M.~C.}\ \bibnamefont {Johnson}}, \bibinfo {author} {\bibfnamefont {L.}~\bibnamefont {Sagunski}}, \bibinfo {author} {\bibfnamefont {M.}~\bibnamefont {Sakellariadou}}, \ and\ \bibinfo {author} {\bibfnamefont {H.}~\bibnamefont {Yang}},\ }\href {\doibase 10.1103/PhysRevLett.127.161101} {\bibfield  {journal} {\bibinfo  {journal} {Phys. Rev. Lett.}\ }\textbf {\bibinfo {volume} {127}},\ \bibinfo {pages} {161101} (\bibinfo {year} {2021})},\ \Eprint {http://arxiv.org/abs/2105.13963} {arXiv:2105.13963 [hep-ph]} \BibitemShut {NoStop}%
\bibitem [{\citenamefont {Balkin}\ \emph {et~al.}(2024)\citenamefont {Balkin}, \citenamefont {Serra}, \citenamefont {Springmann}, \citenamefont {Stelzl},\ and\ \citenamefont {Weiler}}]{Balkin:2022qer}%
  \BibitemOpen
  \bibfield  {author} {\bibinfo {author} {\bibfnamefont {R.}~\bibnamefont {Balkin}}, \bibinfo {author} {\bibfnamefont {J.}~\bibnamefont {Serra}}, \bibinfo {author} {\bibfnamefont {K.}~\bibnamefont {Springmann}}, \bibinfo {author} {\bibfnamefont {S.}~\bibnamefont {Stelzl}}, \ and\ \bibinfo {author} {\bibfnamefont {A.}~\bibnamefont {Weiler}},\ }\href {\doibase 10.1103/PhysRevD.109.095032} {\bibfield  {journal} {\bibinfo  {journal} {Phys. Rev. D}\ }\textbf {\bibinfo {volume} {109}},\ \bibinfo {pages} {095032} (\bibinfo {year} {2024})},\ \Eprint {http://arxiv.org/abs/2211.02661} {arXiv:2211.02661 [hep-ph]} \BibitemShut {NoStop}%
\bibitem [{\citenamefont {G{\'o}mez-Ba{\~n}{\'o}n}\ \emph {et~al.}(2024)\citenamefont {G{\'o}mez-Ba{\~n}{\'o}n}, \citenamefont {Bartnick}, \citenamefont {Springmann},\ and\ \citenamefont {Pons}}]{Gomez-Banon:2024oux}%
  \BibitemOpen
  \bibfield  {author} {\bibinfo {author} {\bibfnamefont {A.}~\bibnamefont {G{\'o}mez-Ba{\~n}{\'o}n}}, \bibinfo {author} {\bibfnamefont {K.}~\bibnamefont {Bartnick}}, \bibinfo {author} {\bibfnamefont {K.}~\bibnamefont {Springmann}}, \ and\ \bibinfo {author} {\bibfnamefont {J.~A.}\ \bibnamefont {Pons}},\ }\href {\doibase 10.1103/PhysRevLett.133.251002} {\bibfield  {journal} {\bibinfo  {journal} {Phys. Rev. Lett.}\ }\textbf {\bibinfo {volume} {133}},\ \bibinfo {pages} {251002} (\bibinfo {year} {2024})},\ \Eprint {http://arxiv.org/abs/2408.07740} {arXiv:2408.07740 [hep-ph]} \BibitemShut {NoStop}%
\bibitem [{\citenamefont {Kumamoto}\ \emph {et~al.}(2025)\citenamefont {Kumamoto}, \citenamefont {Huang}, \citenamefont {Drischler}, \citenamefont {Baryakhtar},\ and\ \citenamefont {Reddy}}]{Kumamoto:2024wjd}%
  \BibitemOpen
  \bibfield  {author} {\bibinfo {author} {\bibfnamefont {M.}~\bibnamefont {Kumamoto}}, \bibinfo {author} {\bibfnamefont {J.}~\bibnamefont {Huang}}, \bibinfo {author} {\bibfnamefont {C.}~\bibnamefont {Drischler}}, \bibinfo {author} {\bibfnamefont {M.}~\bibnamefont {Baryakhtar}}, \ and\ \bibinfo {author} {\bibfnamefont {S.}~\bibnamefont {Reddy}},\ }\href {\doibase 10.1103/5zgm-z4v9} {\bibfield  {journal} {\bibinfo  {journal} {Phys. Rev. D}\ }\textbf {\bibinfo {volume} {112}},\ \bibinfo {pages} {043008} (\bibinfo {year} {2025})},\ \Eprint {http://arxiv.org/abs/2410.21590} {arXiv:2410.21590 [hep-ph]} \BibitemShut {NoStop}%
\bibitem [{\citenamefont {M{\"o}ssbauer}(1958)}]{Mossbauer:1958wsu}%
  \BibitemOpen
  \bibfield  {author} {\bibinfo {author} {\bibfnamefont {R.~L.}\ \bibnamefont {M{\"o}ssbauer}},\ }\href {\doibase 10.1007/BF01344210} {\bibfield  {journal} {\bibinfo  {journal} {Z. Phys.}\ }\textbf {\bibinfo {volume} {151}},\ \bibinfo {pages} {124} (\bibinfo {year} {1958})}\BibitemShut {NoStop}%
\bibitem [{\citenamefont {M\"ossbauer}(1958)}]{Mossbauer2}%
  \BibitemOpen
  \bibfield  {author} {\bibinfo {author} {\bibfnamefont {R.~L.}\ \bibnamefont {M\"ossbauer}},\ }\href@noop {} {\bibfield  {journal} {\bibinfo  {journal} {Naturwissenschaften}\ }\textbf {\bibinfo {volume} {45}},\ \bibinfo {pages} {538} (\bibinfo {year} {1958})}\BibitemShut {NoStop}%
\bibitem [{\citenamefont {Adhikari}\ \emph {et~al.}(2018)\citenamefont {Adhikari}, \citenamefont {Adhikari}, \citenamefont {Souza}, \citenamefont {Carlin}, \citenamefont {Choi}, \citenamefont {Choi}, \citenamefont {Djamal}, \citenamefont {Ezeribe}, \citenamefont {Ha}, \citenamefont {Hahn}, \citenamefont {Hubbard}, \citenamefont {Jeon}, \citenamefont {Jo}, \citenamefont {Joo}, \citenamefont {Kang}, \citenamefont {Kauer}, \citenamefont {Kang}, \citenamefont {Kim}, \citenamefont {Kim}, \citenamefont {Kim}, \citenamefont {Kim}, \citenamefont {Kim}, \citenamefont {Kim}, \citenamefont {Kim}, \citenamefont {Kim}, \citenamefont {Kim}, \citenamefont {Kudryavtsev}, \citenamefont {Lee}, \citenamefont {Lee}, \citenamefont {Lee}, \citenamefont {Lee}, \citenamefont {Leonard}, \citenamefont {Lynch}, \citenamefont {Maruyama}, \citenamefont {Mouton}, \citenamefont {Olsen}, \citenamefont {Park}, \citenamefont {Park}, \citenamefont {Park}, \citenamefont {Park}, \citenamefont {Pettus}, \citenamefont {Prihtiadi}, \citenamefont
  {Ra}, \citenamefont {Rott}, \citenamefont {Scarff}, \citenamefont {Spooner}, \citenamefont {Thompson}, \citenamefont {Yang},\ and\ \citenamefont {Yong}}]{Adhikari_2018}%
  \BibitemOpen
  \bibfield  {author} {\bibinfo {author} {\bibfnamefont {P.}~\bibnamefont {Adhikari}}, \bibinfo {author} {\bibfnamefont {G.}~\bibnamefont {Adhikari}}, \bibinfo {author} {\bibfnamefont {E.~B.~d.}\ \bibnamefont {Souza}}, \bibinfo {author} {\bibfnamefont {N.}~\bibnamefont {Carlin}}, \bibinfo {author} {\bibfnamefont {S.}~\bibnamefont {Choi}}, \bibinfo {author} {\bibfnamefont {W.~Q.}\ \bibnamefont {Choi}}, \bibinfo {author} {\bibfnamefont {M.}~\bibnamefont {Djamal}}, \bibinfo {author} {\bibfnamefont {A.~C.}\ \bibnamefont {Ezeribe}}, \bibinfo {author} {\bibfnamefont {C.}~\bibnamefont {Ha}}, \bibinfo {author} {\bibfnamefont {I.~S.}\ \bibnamefont {Hahn}}, \bibinfo {author} {\bibfnamefont {A.~J.~F.}\ \bibnamefont {Hubbard}}, \bibinfo {author} {\bibfnamefont {E.~J.}\ \bibnamefont {Jeon}}, \bibinfo {author} {\bibfnamefont {J.~H.}\ \bibnamefont {Jo}}, \bibinfo {author} {\bibfnamefont {H.~W.}\ \bibnamefont {Joo}}, \bibinfo {author} {\bibfnamefont {W.~G.}\ \bibnamefont {Kang}}, \bibinfo {author} {\bibfnamefont
  {M.}~\bibnamefont {Kauer}}, \bibinfo {author} {\bibfnamefont {W.~S.}\ \bibnamefont {Kang}}, \bibinfo {author} {\bibfnamefont {B.~H.}\ \bibnamefont {Kim}}, \bibinfo {author} {\bibfnamefont {H.}~\bibnamefont {Kim}}, \bibinfo {author} {\bibfnamefont {H.~J.}\ \bibnamefont {Kim}}, \bibinfo {author} {\bibfnamefont {K.~W.}\ \bibnamefont {Kim}}, \bibinfo {author} {\bibfnamefont {M.~C.}\ \bibnamefont {Kim}}, \bibinfo {author} {\bibfnamefont {N.~Y.}\ \bibnamefont {Kim}}, \bibinfo {author} {\bibfnamefont {S.~K.}\ \bibnamefont {Kim}}, \bibinfo {author} {\bibfnamefont {Y.~D.}\ \bibnamefont {Kim}}, \bibinfo {author} {\bibfnamefont {Y.~H.}\ \bibnamefont {Kim}}, \bibinfo {author} {\bibfnamefont {V.~A.}\ \bibnamefont {Kudryavtsev}}, \bibinfo {author} {\bibfnamefont {H.~S.}\ \bibnamefont {Lee}}, \bibinfo {author} {\bibfnamefont {J.}~\bibnamefont {Lee}}, \bibinfo {author} {\bibfnamefont {J.~Y.}\ \bibnamefont {Lee}}, \bibinfo {author} {\bibfnamefont {M.~H.}\ \bibnamefont {Lee}}, \bibinfo {author} {\bibfnamefont {D.~S.}\
  \bibnamefont {Leonard}}, \bibinfo {author} {\bibfnamefont {W.~A.}\ \bibnamefont {Lynch}}, \bibinfo {author} {\bibfnamefont {R.~H.}\ \bibnamefont {Maruyama}}, \bibinfo {author} {\bibfnamefont {F.}~\bibnamefont {Mouton}}, \bibinfo {author} {\bibfnamefont {S.~L.}\ \bibnamefont {Olsen}}, \bibinfo {author} {\bibfnamefont {H.~K.}\ \bibnamefont {Park}}, \bibinfo {author} {\bibfnamefont {H.~S.}\ \bibnamefont {Park}}, \bibinfo {author} {\bibfnamefont {J.~S.}\ \bibnamefont {Park}}, \bibinfo {author} {\bibfnamefont {K.~S.}\ \bibnamefont {Park}}, \bibinfo {author} {\bibfnamefont {W.}~\bibnamefont {Pettus}}, \bibinfo {author} {\bibfnamefont {H.}~\bibnamefont {Prihtiadi}}, \bibinfo {author} {\bibfnamefont {S.}~\bibnamefont {Ra}}, \bibinfo {author} {\bibfnamefont {C.}~\bibnamefont {Rott}}, \bibinfo {author} {\bibfnamefont {A.}~\bibnamefont {Scarff}}, \bibinfo {author} {\bibfnamefont {N.~J.~C.}\ \bibnamefont {Spooner}}, \bibinfo {author} {\bibfnamefont {W.~G.}\ \bibnamefont {Thompson}}, \bibinfo {author} {\bibfnamefont
  {L.}~\bibnamefont {Yang}}, \ and\ \bibinfo {author} {\bibfnamefont {S.~H.}\ \bibnamefont {Yong}},\ }\href {\doibase 10.1140/epjc/s10052-018-5970-2} {\bibfield  {journal} {\bibinfo  {journal} {The European Physical Journal C}\ }\textbf {\bibinfo {volume} {78}} (\bibinfo {year} {2018}),\ 10.1140/epjc/s10052-018-5970-2}\BibitemShut {NoStop}%
\bibitem [{\citenamefont {L.R.}(1962)}]{LR1962691}%
  \BibitemOpen
  \bibfield  {author} {\bibinfo {author} {\bibnamefont {L.R.}},\ }\href {\doibase https://doi.org/10.1016/0029-5582(62)90301-2} {\bibfield  {journal} {\bibinfo  {journal} {Nuclear Physics}\ }\textbf {\bibinfo {volume} {37}},\ \bibinfo {pages} {691} (\bibinfo {year} {1962})}\BibitemShut {NoStop}%
\bibitem [{\citenamefont {Wildner}\ and\ \citenamefont {Gonser}(1979)}]{wildner:jpa-00218534}%
  \BibitemOpen
  \bibfield  {author} {\bibinfo {author} {\bibfnamefont {W.}~\bibnamefont {Wildner}}\ and\ \bibinfo {author} {\bibfnamefont {U.}~\bibnamefont {Gonser}},\ }\href {\doibase 10.1051/jphyscol:1979216} {\bibfield  {journal} {\bibinfo  {journal} {{Journal de Physique Colloques}}\ }\textbf {\bibinfo {volume} {40}},\ \bibinfo {pages} {C2} (\bibinfo {year} {1979})}\BibitemShut {NoStop}%
\bibitem [{\citenamefont {Taylor}\ and\ \citenamefont {Hoy}(1988)}]{Taylor:1988}%
  \BibitemOpen
  \bibfield  {author} {\bibinfo {author} {\bibfnamefont {R.~D.}\ \bibnamefont {Taylor}}\ and\ \bibinfo {author} {\bibfnamefont {G.~R.}\ \bibnamefont {Hoy}},\ }in\ \href {\doibase 10.1117/12.943893} {\emph {\bibinfo {booktitle} {Short and Ultrashort Wavelength Lasers}}},\ Vol.\ \bibinfo {volume} {875},\ \bibinfo {editor} {edited by\ \bibinfo {editor} {\bibfnamefont {R.~C.~R.}\ \bibnamefont {Jones}}}\ (\bibinfo  {publisher} {SPIE},\ \bibinfo {year} {1988})\ pp.\ \bibinfo {pages} {126--135}\BibitemShut {NoStop}%
\bibitem [{\citenamefont {Hoy}\ \emph {et~al.}(1990)\citenamefont {Hoy}, \citenamefont {Rezaie-Serej},\ and\ \citenamefont {Taylor}}]{Hoy:1990}%
  \BibitemOpen
  \bibfield  {author} {\bibinfo {author} {\bibfnamefont {G.~R.}\ \bibnamefont {Hoy}}, \bibinfo {author} {\bibfnamefont {S.}~\bibnamefont {Rezaie-Serej}}, \ and\ \bibinfo {author} {\bibfnamefont {R.~D.}\ \bibnamefont {Taylor}},\ }\href {\doibase 10.1134/S0021364009190011} {\bibfield  {journal} {\bibinfo  {journal} {Hyperfine Interactions}\ }\textbf {\bibinfo {volume} {58}},\ \bibinfo {pages} {2513} (\bibinfo {year} {1990})}\BibitemShut {NoStop}%
\bibitem [{\citenamefont {Alpatov}\ \emph {et~al.}(1996)\citenamefont {Alpatov}, \citenamefont {Bayukov}, \citenamefont {Davydov}, \citenamefont {Isaev}, \citenamefont {Kartashov}, \citenamefont {Korotkov},\ and\ \citenamefont {Samoilov}}]{Alpatov:1996bd}%
  \BibitemOpen
  \bibfield  {author} {\bibinfo {author} {\bibfnamefont {V.~G.}\ \bibnamefont {Alpatov}}, \bibinfo {author} {\bibfnamefont {Y.~D.}\ \bibnamefont {Bayukov}}, \bibinfo {author} {\bibfnamefont {A.~V.}\ \bibnamefont {Davydov}}, \bibinfo {author} {\bibfnamefont {Y.~N.}\ \bibnamefont {Isaev}}, \bibinfo {author} {\bibfnamefont {G.~R.}\ \bibnamefont {Kartashov}}, \bibinfo {author} {\bibfnamefont {M.~M.}\ \bibnamefont {Korotkov}}, \ and\ \bibinfo {author} {\bibfnamefont {V.~M.}\ \bibnamefont {Samoilov}},\ }\href@noop {} {\  (\bibinfo {year} {1996})}\BibitemShut {NoStop}%
\bibitem [{\citenamefont {Alpatov}\ \emph {et~al.}(2007)\citenamefont {Alpatov}, \citenamefont {Bayukov}, \citenamefont {Davydov}, \citenamefont {Isaev}, \citenamefont {Kartashov}, \citenamefont {Korotkov},\ and\ \citenamefont {Migachev}}]{Alpatov:2007LF}%
  \BibitemOpen
  \bibfield  {author} {\bibinfo {author} {\bibfnamefont {V.~G.}\ \bibnamefont {Alpatov}}, \bibinfo {author} {\bibfnamefont {Y.~D.}\ \bibnamefont {Bayukov}}, \bibinfo {author} {\bibfnamefont {A.~V.}\ \bibnamefont {Davydov}}, \bibinfo {author} {\bibfnamefont {Y.~N.}\ \bibnamefont {Isaev}}, \bibinfo {author} {\bibfnamefont {G.~R.}\ \bibnamefont {Kartashov}}, \bibinfo {author} {\bibfnamefont {M.~M.}\ \bibnamefont {Korotkov}}, \ and\ \bibinfo {author} {\bibfnamefont {V.~V.}\ \bibnamefont {Migachev}},\ }\href {\doibase 10.1134/S1054660X07080087} {\bibfield  {journal} {\bibinfo  {journal} {Laser Physics}\ }\textbf {\bibinfo {volume} {17}},\ \bibinfo {pages} {1067} (\bibinfo {year} {2007})}\BibitemShut {NoStop}%
\bibitem [{\citenamefont {Bayukov}\ \emph {et~al.}(2009)\citenamefont {Bayukov}, \citenamefont {Davydov}, \citenamefont {Isaev}, \citenamefont {Kartashov}, \citenamefont {Korotkov},\ and\ \citenamefont {Migachev}}]{Bayukov2009}%
  \BibitemOpen
  \bibfield  {author} {\bibinfo {author} {\bibfnamefont {Y.~D.}\ \bibnamefont {Bayukov}}, \bibinfo {author} {\bibfnamefont {A.~V.}\ \bibnamefont {Davydov}}, \bibinfo {author} {\bibfnamefont {Y.~N.}\ \bibnamefont {Isaev}}, \bibinfo {author} {\bibfnamefont {G.~R.}\ \bibnamefont {Kartashov}}, \bibinfo {author} {\bibfnamefont {M.~M.}\ \bibnamefont {Korotkov}}, \ and\ \bibinfo {author} {\bibfnamefont {V.~V.}\ \bibnamefont {Migachev}},\ }\href {\doibase 10.1134/S0021364009190011} {\bibfield  {journal} {\bibinfo  {journal} {JETP Letters}\ }\textbf {\bibinfo {volume} {90}},\ \bibinfo {pages} {499} (\bibinfo {year} {2009})}\BibitemShut {NoStop}%
\bibitem [{\citenamefont {Alpatov}\ \emph {et~al.}(2008)\citenamefont {Alpatov}, \citenamefont {Bayukov}, \citenamefont {Davydov}, \citenamefont {Isaev}, \citenamefont {Kartashov}, \citenamefont {Korotkov},\ and\ \citenamefont {Migachev}}]{Alpatov2008}%
  \BibitemOpen
  \bibfield  {author} {\bibinfo {author} {\bibfnamefont {V.~G.}\ \bibnamefont {Alpatov}}, \bibinfo {author} {\bibfnamefont {Y.~D.}\ \bibnamefont {Bayukov}}, \bibinfo {author} {\bibfnamefont {A.~V.}\ \bibnamefont {Davydov}}, \bibinfo {author} {\bibfnamefont {Y.~N.}\ \bibnamefont {Isaev}}, \bibinfo {author} {\bibfnamefont {G.~R.}\ \bibnamefont {Kartashov}}, \bibinfo {author} {\bibfnamefont {M.~M.}\ \bibnamefont {Korotkov}}, \ and\ \bibinfo {author} {\bibfnamefont {V.~V.}\ \bibnamefont {Migachev}},\ }\href {\doibase 10.1134/S1063778808070053} {\bibfield  {journal} {\bibinfo  {journal} {Physics of Atomic Nuclei}\ }\textbf {\bibinfo {volume} {71}},\ \bibinfo {pages} {1156} (\bibinfo {year} {2008})}\BibitemShut {NoStop}%
\bibitem [{\citenamefont {Ubaldi}(2010)}]{Ubaldi:2008nf}%
  \BibitemOpen
  \bibfield  {author} {\bibinfo {author} {\bibfnamefont {L.}~\bibnamefont {Ubaldi}},\ }\href {\doibase 10.1103/PhysRevD.81.025011} {\bibfield  {journal} {\bibinfo  {journal} {Phys. Rev. D}\ }\textbf {\bibinfo {volume} {81}},\ \bibinfo {pages} {025011} (\bibinfo {year} {2010})},\ \Eprint {http://arxiv.org/abs/0811.1599} {arXiv:0811.1599 [hep-ph]} \BibitemShut {NoStop}%
\bibitem [{\citenamefont {Damour}\ and\ \citenamefont {Donoghue}(2008)}]{Damour:2007uv}%
  \BibitemOpen
  \bibfield  {author} {\bibinfo {author} {\bibfnamefont {T.}~\bibnamefont {Damour}}\ and\ \bibinfo {author} {\bibfnamefont {J.~F.}\ \bibnamefont {Donoghue}},\ }\href {\doibase 10.1103/PhysRevD.78.014014} {\bibfield  {journal} {\bibinfo  {journal} {Phys. Rev. D}\ }\textbf {\bibinfo {volume} {78}},\ \bibinfo {pages} {014014} (\bibinfo {year} {2008})},\ \Eprint {http://arxiv.org/abs/0712.2968} {arXiv:0712.2968 [hep-ph]} \BibitemShut {NoStop}%
\bibitem [{\citenamefont {Hohenberg}\ and\ \citenamefont {Kohn}(1964)}]{Hohenberg1964PhysRev.136.B864}%
  \BibitemOpen
  \bibfield  {author} {\bibinfo {author} {\bibfnamefont {P.}~\bibnamefont {Hohenberg}}\ and\ \bibinfo {author} {\bibfnamefont {W.}~\bibnamefont {Kohn}},\ }\href {\doibase 10.1103/PhysRev.136.B864} {\bibfield  {journal} {\bibinfo  {journal} {Phys. Rev.}\ }\textbf {\bibinfo {volume} {136}},\ \bibinfo {pages} {B864} (\bibinfo {year} {1964})}\BibitemShut {NoStop}%
\bibitem [{\citenamefont {Serot}\ and\ \citenamefont {Walecka}(1997)}]{Serot1997IJMPE6.515}%
  \BibitemOpen
  \bibfield  {author} {\bibinfo {author} {\bibfnamefont {B.~D.}\ \bibnamefont {Serot}}\ and\ \bibinfo {author} {\bibfnamefont {J.~D.}\ \bibnamefont {Walecka}},\ }\href@noop {} {\bibfield  {journal} {\bibinfo  {journal} {Int. Jour. Mod. Phys. E}\ }\textbf {\bibinfo {volume} {6}},\ \bibinfo {pages} {515} (\bibinfo {year} {1997})}\BibitemShut {NoStop}%
\bibitem [{\citenamefont {Ring}(1996)}]{Ring1996PPNP.37.193}%
  \BibitemOpen
  \bibfield  {author} {\bibinfo {author} {\bibfnamefont {P.}~\bibnamefont {Ring}},\ }\href {\doibase https://doi.org/10.1016/0146-6410(96)00054-3} {\bibfield  {journal} {\bibinfo  {journal} {Prog. Part. Nucl. Phys.}\ }\textbf {\bibinfo {volume} {37}},\ \bibinfo {pages} {193} (\bibinfo {year} {1996})}\BibitemShut {NoStop}%
\bibitem [{\citenamefont {Zhao}\ \emph {et~al.}(2010)\citenamefont {Zhao}, \citenamefont {Li}, \citenamefont {Yao},\ and\ \citenamefont {Meng}}]{Zhao2012PhysRevC.82.054319}%
  \BibitemOpen
  \bibfield  {author} {\bibinfo {author} {\bibfnamefont {P.~W.}\ \bibnamefont {Zhao}}, \bibinfo {author} {\bibfnamefont {Z.~P.}\ \bibnamefont {Li}}, \bibinfo {author} {\bibfnamefont {J.~M.}\ \bibnamefont {Yao}}, \ and\ \bibinfo {author} {\bibfnamefont {J.}~\bibnamefont {Meng}},\ }\href {\doibase 10.1103/PhysRevC.82.054319} {\bibfield  {journal} {\bibinfo  {journal} {Phys. Rev. C}\ }\textbf {\bibinfo {volume} {82}},\ \bibinfo {pages} {054319} (\bibinfo {year} {2010})}\BibitemShut {NoStop}%
\bibitem [{\citenamefont {Yang}\ \emph {et~al.}(2021)\citenamefont {Yang}, \citenamefont {Wang}, \citenamefont {Zhao},\ and\ \citenamefont {Li}}]{Yang2021PhysRevC.104.054312}%
  \BibitemOpen
  \bibfield  {author} {\bibinfo {author} {\bibfnamefont {Y.~L.}\ \bibnamefont {Yang}}, \bibinfo {author} {\bibfnamefont {Y.~K.}\ \bibnamefont {Wang}}, \bibinfo {author} {\bibfnamefont {P.~W.}\ \bibnamefont {Zhao}}, \ and\ \bibinfo {author} {\bibfnamefont {Z.~P.}\ \bibnamefont {Li}},\ }\href {\doibase 10.1103/PhysRevC.104.054312} {\bibfield  {journal} {\bibinfo  {journal} {Phys. Rev. C}\ }\textbf {\bibinfo {volume} {104}},\ \bibinfo {pages} {054312} (\bibinfo {year} {2021})}\BibitemShut {NoStop}%
\bibitem [{\citenamefont {Zhang}\ \emph {et~al.}(2022)\citenamefont {Zhang}, \citenamefont {Cheoun}, \citenamefont {Choi}, \citenamefont {Chong}, \citenamefont {Dong}, \citenamefont {Dong}, \citenamefont {Du}, \citenamefont {Geng}, \citenamefont {Ha}, \citenamefont {He} \emph {et~al.}}]{Zhang2022ADNDT144.101488}%
  \BibitemOpen
  \bibfield  {author} {\bibinfo {author} {\bibfnamefont {K.}~\bibnamefont {Zhang}}, \bibinfo {author} {\bibfnamefont {M.-K.}\ \bibnamefont {Cheoun}}, \bibinfo {author} {\bibfnamefont {Y.-B.}\ \bibnamefont {Choi}}, \bibinfo {author} {\bibfnamefont {P.~S.}\ \bibnamefont {Chong}}, \bibinfo {author} {\bibfnamefont {J.}~\bibnamefont {Dong}}, \bibinfo {author} {\bibfnamefont {Z.}~\bibnamefont {Dong}}, \bibinfo {author} {\bibfnamefont {X.}~\bibnamefont {Du}}, \bibinfo {author} {\bibfnamefont {L.}~\bibnamefont {Geng}}, \bibinfo {author} {\bibfnamefont {E.}~\bibnamefont {Ha}}, \bibinfo {author} {\bibfnamefont {X.-T.}\ \bibnamefont {He}},  \emph {et~al.},\ }\href@noop {} {\bibfield  {journal} {\bibinfo  {journal} {Atom. Data Nucl. Data Tabl.}\ }\textbf {\bibinfo {volume} {144}},\ \bibinfo {pages} {101488} (\bibinfo {year} {2022})}\BibitemShut {NoStop}%
\bibitem [{\citenamefont {Guo}\ \emph {et~al.}(2024)\citenamefont {Guo}, \citenamefont {Cao}, \citenamefont {Chen}, \citenamefont {Chen}, \citenamefont {Cheoun}, \citenamefont {Choi}, \citenamefont {Lam}, \citenamefont {Deng}, \citenamefont {Dong}, \citenamefont {Du} \emph {et~al.}}]{Guo2024ADNDT158.101661}%
  \BibitemOpen
  \bibfield  {author} {\bibinfo {author} {\bibfnamefont {P.}~\bibnamefont {Guo}}, \bibinfo {author} {\bibfnamefont {X.}~\bibnamefont {Cao}}, \bibinfo {author} {\bibfnamefont {K.}~\bibnamefont {Chen}}, \bibinfo {author} {\bibfnamefont {Z.}~\bibnamefont {Chen}}, \bibinfo {author} {\bibfnamefont {M.-K.}\ \bibnamefont {Cheoun}}, \bibinfo {author} {\bibfnamefont {Y.-B.}\ \bibnamefont {Choi}}, \bibinfo {author} {\bibfnamefont {P.~C.}\ \bibnamefont {Lam}}, \bibinfo {author} {\bibfnamefont {W.}~\bibnamefont {Deng}}, \bibinfo {author} {\bibfnamefont {J.}~\bibnamefont {Dong}}, \bibinfo {author} {\bibfnamefont {P.}~\bibnamefont {Du}},  \emph {et~al.},\ }\href@noop {} {\bibfield  {journal} {\bibinfo  {journal} {Atom. Data Nucl. Data Tabl.}\ }\textbf {\bibinfo {volume} {158}},\ \bibinfo {pages} {101661} (\bibinfo {year} {2024})}\BibitemShut {NoStop}%
\bibitem [{\citenamefont {Zhao}\ \emph {et~al.}(2011)\citenamefont {Zhao}, \citenamefont {Peng}, \citenamefont {Liang}, \citenamefont {Ring},\ and\ \citenamefont {Meng}}]{Zhao2011PhysRevLett.107.122501}%
  \BibitemOpen
  \bibfield  {author} {\bibinfo {author} {\bibfnamefont {P.~W.}\ \bibnamefont {Zhao}}, \bibinfo {author} {\bibfnamefont {J.}~\bibnamefont {Peng}}, \bibinfo {author} {\bibfnamefont {H.~Z.}\ \bibnamefont {Liang}}, \bibinfo {author} {\bibfnamefont {P.}~\bibnamefont {Ring}}, \ and\ \bibinfo {author} {\bibfnamefont {J.}~\bibnamefont {Meng}},\ }\href {\doibase 10.1103/PhysRevLett.107.122501} {\bibfield  {journal} {\bibinfo  {journal} {Phys. Rev. Lett.}\ }\textbf {\bibinfo {volume} {107}},\ \bibinfo {pages} {122501} (\bibinfo {year} {2011})}\BibitemShut {NoStop}%
\bibitem [{\citenamefont {Zhao}(2017)}]{Zhao2017PLB773.1}%
  \BibitemOpen
  \bibfield  {author} {\bibinfo {author} {\bibfnamefont {P.~W.}\ \bibnamefont {Zhao}},\ }\href@noop {} {\bibfield  {journal} {\bibinfo  {journal} {Phys. Lett. B}\ }\textbf {\bibinfo {volume} {773}},\ \bibinfo {pages} {1} (\bibinfo {year} {2017})}\BibitemShut {NoStop}%
\bibitem [{\citenamefont {Wang}\ \emph {et~al.}(2024{\natexlab{a}})\citenamefont {Wang}, \citenamefont {Zhao},\ and\ \citenamefont {Meng}}]{Wang2024PLB848.138346}%
  \BibitemOpen
  \bibfield  {author} {\bibinfo {author} {\bibfnamefont {Y.~K.}\ \bibnamefont {Wang}}, \bibinfo {author} {\bibfnamefont {P.~W.}\ \bibnamefont {Zhao}}, \ and\ \bibinfo {author} {\bibfnamefont {J.}~\bibnamefont {Meng}},\ }\href@noop {} {\bibfield  {journal} {\bibinfo  {journal} {Phys. Lett. B}\ }\textbf {\bibinfo {volume} {848}},\ \bibinfo {pages} {138346} (\bibinfo {year} {2024}{\natexlab{a}})}\BibitemShut {NoStop}%
\bibitem [{\citenamefont {Ren}\ \emph {et~al.}(2022)\citenamefont {Ren}, \citenamefont {Vretenar}, \citenamefont {Nik\ifmmode \check{s}\else \v{s}\fi{}i\ifmmode~\acute{c}\else \'{c}\fi{}}, \citenamefont {Zhao}, \citenamefont {Zhao},\ and\ \citenamefont {Meng}}]{Ren2022PhysRevLett.128.172501}%
  \BibitemOpen
  \bibfield  {author} {\bibinfo {author} {\bibfnamefont {Z.~X.}\ \bibnamefont {Ren}}, \bibinfo {author} {\bibfnamefont {D.}~\bibnamefont {Vretenar}}, \bibinfo {author} {\bibfnamefont {T.}~\bibnamefont {Nik\ifmmode \check{s}\else \v{s}\fi{}i\ifmmode~\acute{c}\else \'{c}\fi{}}}, \bibinfo {author} {\bibfnamefont {P.~W.}\ \bibnamefont {Zhao}}, \bibinfo {author} {\bibfnamefont {J.}~\bibnamefont {Zhao}}, \ and\ \bibinfo {author} {\bibfnamefont {J.}~\bibnamefont {Meng}},\ }\href {\doibase 10.1103/PhysRevLett.128.172501} {\bibfield  {journal} {\bibinfo  {journal} {Phys. Rev. Lett.}\ }\textbf {\bibinfo {volume} {128}},\ \bibinfo {pages} {172501} (\bibinfo {year} {2022})}\BibitemShut {NoStop}%
\bibitem [{\citenamefont {Zhang}\ \emph {et~al.}(2024)\citenamefont {Zhang}, \citenamefont {Li}, \citenamefont {Vretenar}, \citenamefont {Nik\ifmmode \check{s}\else \v{s}\fi{}i\ifmmode~\acute{c}\else \'{c}\fi{}}, \citenamefont {Ren}, \citenamefont {Zhao},\ and\ \citenamefont {Meng}}]{Zhang2024PhysRevC.109.024316}%
  \BibitemOpen
  \bibfield  {author} {\bibinfo {author} {\bibfnamefont {D.~D.}\ \bibnamefont {Zhang}}, \bibinfo {author} {\bibfnamefont {B.}~\bibnamefont {Li}}, \bibinfo {author} {\bibfnamefont {D.}~\bibnamefont {Vretenar}}, \bibinfo {author} {\bibfnamefont {T.}~\bibnamefont {Nik\ifmmode \check{s}\else \v{s}\fi{}i\ifmmode~\acute{c}\else \'{c}\fi{}}}, \bibinfo {author} {\bibfnamefont {Z.~X.}\ \bibnamefont {Ren}}, \bibinfo {author} {\bibfnamefont {P.~W.}\ \bibnamefont {Zhao}}, \ and\ \bibinfo {author} {\bibfnamefont {J.}~\bibnamefont {Meng}},\ }\href {\doibase 10.1103/PhysRevC.109.024316} {\bibfield  {journal} {\bibinfo  {journal} {Phys. Rev. C}\ }\textbf {\bibinfo {volume} {109}},\ \bibinfo {pages} {024316} (\bibinfo {year} {2024})}\BibitemShut {NoStop}%
\bibitem [{\citenamefont {Li}\ \emph {et~al.}(2023)\citenamefont {Li}, \citenamefont {Vretenar}, \citenamefont {Ren}, \citenamefont {Nik\ifmmode \check{s}\else \v{s}\fi{}i\ifmmode~\acute{c}\else \'{c}\fi{}}, \citenamefont {Zhao}, \citenamefont {Zhao},\ and\ \citenamefont {Meng}}]{Li2023PhysRevC.107.014303}%
  \BibitemOpen
  \bibfield  {author} {\bibinfo {author} {\bibfnamefont {B.}~\bibnamefont {Li}}, \bibinfo {author} {\bibfnamefont {D.}~\bibnamefont {Vretenar}}, \bibinfo {author} {\bibfnamefont {Z.~X.}\ \bibnamefont {Ren}}, \bibinfo {author} {\bibfnamefont {T.}~\bibnamefont {Nik\ifmmode \check{s}\else \v{s}\fi{}i\ifmmode~\acute{c}\else \'{c}\fi{}}}, \bibinfo {author} {\bibfnamefont {J.}~\bibnamefont {Zhao}}, \bibinfo {author} {\bibfnamefont {P.~W.}\ \bibnamefont {Zhao}}, \ and\ \bibinfo {author} {\bibfnamefont {J.}~\bibnamefont {Meng}},\ }\href {\doibase 10.1103/PhysRevC.107.014303} {\bibfield  {journal} {\bibinfo  {journal} {Phys. Rev. C}\ }\textbf {\bibinfo {volume} {107}},\ \bibinfo {pages} {014303} (\bibinfo {year} {2023})}\BibitemShut {NoStop}%
\bibitem [{\citenamefont {Wang}\ \emph {et~al.}(2024{\natexlab{b}})\citenamefont {Wang}, \citenamefont {Zhao},\ and\ \citenamefont {Meng}}]{Wang2024Sci.Bull.2017}%
  \BibitemOpen
  \bibfield  {author} {\bibinfo {author} {\bibfnamefont {Y.}~\bibnamefont {Wang}}, \bibinfo {author} {\bibfnamefont {P.}~\bibnamefont {Zhao}}, \ and\ \bibinfo {author} {\bibfnamefont {J.}~\bibnamefont {Meng}},\ }\href {\doibase https://doi.org/10.1016/j.scib.2024.04.071} {\bibfield  {journal} {\bibinfo  {journal} {Sci. Bull.}\ }\textbf {\bibinfo {volume} {69}},\ \bibinfo {pages} {2017} (\bibinfo {year} {2024}{\natexlab{b}})}\BibitemShut {NoStop}%
\bibitem [{\citenamefont {Wang}\ \emph {et~al.}(2024{\natexlab{c}})\citenamefont {Wang}, \citenamefont {Zhao},\ and\ \citenamefont {Meng}}]{Wang2024PLB855.138796}%
  \BibitemOpen
  \bibfield  {author} {\bibinfo {author} {\bibfnamefont {Y.}~\bibnamefont {Wang}}, \bibinfo {author} {\bibfnamefont {P.}~\bibnamefont {Zhao}}, \ and\ \bibinfo {author} {\bibfnamefont {J.}~\bibnamefont {Meng}},\ }\href@noop {} {\bibfield  {journal} {\bibinfo  {journal} {Physics Letters B}\ }\textbf {\bibinfo {volume} {855}},\ \bibinfo {pages} {138796} (\bibinfo {year} {2024}{\natexlab{c}})}\BibitemShut {NoStop}%
\bibitem [{\citenamefont {Wang}\ \emph {et~al.}(2021)\citenamefont {Wang}, \citenamefont {Huang}, \citenamefont {Kondev}, \citenamefont {Audi},\ and\ \citenamefont {Naimi}}]{Wang2021AME}%
  \BibitemOpen
  \bibfield  {author} {\bibinfo {author} {\bibfnamefont {M.}~\bibnamefont {Wang}}, \bibinfo {author} {\bibfnamefont {W.~J.}\ \bibnamefont {Huang}}, \bibinfo {author} {\bibfnamefont {F.~G.}\ \bibnamefont {Kondev}}, \bibinfo {author} {\bibfnamefont {G.}~\bibnamefont {Audi}}, \ and\ \bibinfo {author} {\bibfnamefont {S.}~\bibnamefont {Naimi}},\ }\href@noop {} {\bibfield  {journal} {\bibinfo  {journal} {Chin. Phys. C}\ }\textbf {\bibinfo {volume} {45}},\ \bibinfo {pages} {030003} (\bibinfo {year} {2021})}\BibitemShut {NoStop}%
\bibitem [{\citenamefont {Leutz}\ \emph {et~al.}(1965)\citenamefont {Leutz}, \citenamefont {Schneckenberger},\ and\ \citenamefont {Wenninger}}]{LEUTZ1965263}%
  \BibitemOpen
  \bibfield  {author} {\bibinfo {author} {\bibfnamefont {H.}~\bibnamefont {Leutz}}, \bibinfo {author} {\bibfnamefont {K.}~\bibnamefont {Schneckenberger}}, \ and\ \bibinfo {author} {\bibfnamefont {H.}~\bibnamefont {Wenninger}},\ }\href {\doibase https://doi.org/10.1016/0029-5582(65)90342-1} {\bibfield  {journal} {\bibinfo  {journal} {Nuclear Physics}\ }\textbf {\bibinfo {volume} {63}},\ \bibinfo {pages} {263} (\bibinfo {year} {1965})}\BibitemShut {NoStop}%
\bibitem [{\citenamefont {{Idaho National Laboratory}}(1999)}]{INL_Cd109}%
  \BibitemOpen
  \bibfield  {author} {\bibinfo {author} {\bibnamefont {{Idaho National Laboratory}}},\ }\href {https://gammaray.inl.gov/SiteAssets/catalogs/ge/pdf/cd109.pdf} {\emph {\bibinfo {title} {{“\^{}109Cd (462 day) Decay Scheme: Gamma-Ray Energies and Intensities”}}}},\ \bibinfo {type} {Spectrum Catalog}\ (\bibinfo  {institution} {Idaho National Laboratory (INL)},\ \bibinfo {year} {1999})\BibitemShut {NoStop}%
\bibitem [{\citenamefont {Gao}\ \emph {et~al.}(2024)\citenamefont {Gao}, \citenamefont {Zhang},\ and\ \citenamefont {Xu}}]{Gao:2023ggo}%
  \BibitemOpen
  \bibfield  {author} {\bibinfo {author} {\bibfnamefont {Y.}~\bibnamefont {Gao}}, \bibinfo {author} {\bibfnamefont {H.}~\bibnamefont {Zhang}}, \ and\ \bibinfo {author} {\bibfnamefont {W.}~\bibnamefont {Xu}},\ }\href {\doibase 10.1016/j.scib.2024.07.038} {\bibfield  {journal} {\bibinfo  {journal} {Sci. Bull.}\ }\textbf {\bibinfo {volume} {69}},\ \bibinfo {pages} {2795} (\bibinfo {year} {2024})},\ \Eprint {http://arxiv.org/abs/2310.06607} {arXiv:2310.06607 [gr-qc]} \BibitemShut {NoStop}%
\bibitem [{\citenamefont {Roussy}\ \emph {et~al.}(2021)\citenamefont {Roussy} \emph {et~al.}}]{Roussy:2020ily}%
  \BibitemOpen
  \bibfield  {author} {\bibinfo {author} {\bibfnamefont {T.~S.}\ \bibnamefont {Roussy}} \emph {et~al.},\ }\href {\doibase 10.1103/PhysRevLett.126.171301} {\bibfield  {journal} {\bibinfo  {journal} {Phys. Rev. Lett.}\ }\textbf {\bibinfo {volume} {126}},\ \bibinfo {pages} {171301} (\bibinfo {year} {2021})},\ \Eprint {http://arxiv.org/abs/2006.15787} {arXiv:2006.15787 [hep-ph]} \BibitemShut {NoStop}%
\bibitem [{\citenamefont {Madge}\ \emph {et~al.}(2024)\citenamefont {Madge}, \citenamefont {Perez},\ and\ \citenamefont {Meir}}]{Madge:2024aot}%
  \BibitemOpen
  \bibfield  {author} {\bibinfo {author} {\bibfnamefont {E.}~\bibnamefont {Madge}}, \bibinfo {author} {\bibfnamefont {G.}~\bibnamefont {Perez}}, \ and\ \bibinfo {author} {\bibfnamefont {Z.}~\bibnamefont {Meir}},\ }\href {\doibase 10.1103/PhysRevD.110.015008} {\bibfield  {journal} {\bibinfo  {journal} {Phys. Rev. D}\ }\textbf {\bibinfo {volume} {110}},\ \bibinfo {pages} {015008} (\bibinfo {year} {2024})},\ \Eprint {http://arxiv.org/abs/2404.00616} {arXiv:2404.00616 [physics.atom-ph]} \BibitemShut {NoStop}%
\bibitem [{\citenamefont {Banerjee}\ \emph {et~al.}(2023)\citenamefont {Banerjee}, \citenamefont {Budker}, \citenamefont {Filzinger}, \citenamefont {Huntemann}, \citenamefont {Paz}, \citenamefont {Perez}, \citenamefont {Porsev},\ and\ \citenamefont {Safronova}}]{Banerjee:2023bjc}%
  \BibitemOpen
  \bibfield  {author} {\bibinfo {author} {\bibfnamefont {A.}~\bibnamefont {Banerjee}}, \bibinfo {author} {\bibfnamefont {D.}~\bibnamefont {Budker}}, \bibinfo {author} {\bibfnamefont {M.}~\bibnamefont {Filzinger}}, \bibinfo {author} {\bibfnamefont {N.}~\bibnamefont {Huntemann}}, \bibinfo {author} {\bibfnamefont {G.}~\bibnamefont {Paz}}, \bibinfo {author} {\bibfnamefont {G.}~\bibnamefont {Perez}}, \bibinfo {author} {\bibfnamefont {S.}~\bibnamefont {Porsev}}, \ and\ \bibinfo {author} {\bibfnamefont {M.}~\bibnamefont {Safronova}},\ }\href@noop {} {\  (\bibinfo {year} {2023})},\ \Eprint {http://arxiv.org/abs/2301.10784} {arXiv:2301.10784 [hep-ph]} \BibitemShut {NoStop}%
\bibitem [{\citenamefont {Mehta}\ \emph {et~al.}(2020)\citenamefont {Mehta}, \citenamefont {Demirtas}, \citenamefont {Long}, \citenamefont {Marsh}, \citenamefont {Mcallister},\ and\ \citenamefont {Stott}}]{Mehta:2020kwu}%
  \BibitemOpen
  \bibfield  {author} {\bibinfo {author} {\bibfnamefont {V.~M.}\ \bibnamefont {Mehta}}, \bibinfo {author} {\bibfnamefont {M.}~\bibnamefont {Demirtas}}, \bibinfo {author} {\bibfnamefont {C.}~\bibnamefont {Long}}, \bibinfo {author} {\bibfnamefont {D.~J.~E.}\ \bibnamefont {Marsh}}, \bibinfo {author} {\bibfnamefont {L.}~\bibnamefont {Mcallister}}, \ and\ \bibinfo {author} {\bibfnamefont {M.~J.}\ \bibnamefont {Stott}},\ }\href@noop {} {\  (\bibinfo {year} {2020})},\ \Eprint {http://arxiv.org/abs/2011.08693} {arXiv:2011.08693 [hep-th]} \BibitemShut {NoStop}%
\bibitem [{\citenamefont {Baryakhtar}\ \emph {et~al.}(2021)\citenamefont {Baryakhtar}, \citenamefont {Galanis}, \citenamefont {Lasenby},\ and\ \citenamefont {Simon}}]{Baryakhtar:2020gao}%
  \BibitemOpen
  \bibfield  {author} {\bibinfo {author} {\bibfnamefont {M.}~\bibnamefont {Baryakhtar}}, \bibinfo {author} {\bibfnamefont {M.}~\bibnamefont {Galanis}}, \bibinfo {author} {\bibfnamefont {R.}~\bibnamefont {Lasenby}}, \ and\ \bibinfo {author} {\bibfnamefont {O.}~\bibnamefont {Simon}},\ }\href {\doibase 10.1103/PhysRevD.103.095019} {\bibfield  {journal} {\bibinfo  {journal} {Phys. Rev. D}\ }\textbf {\bibinfo {volume} {103}},\ \bibinfo {pages} {095019} (\bibinfo {year} {2021})},\ \Eprint {http://arxiv.org/abs/2011.11646} {arXiv:2011.11646 [hep-ph]} \BibitemShut {NoStop}%
\bibitem [{\citenamefont {{\"U}nal}\ \emph {et~al.}(2021)\citenamefont {{\"U}nal}, \citenamefont {Pacucci},\ and\ \citenamefont {Loeb}}]{Unal:2020jiy}%
  \BibitemOpen
  \bibfield  {author} {\bibinfo {author} {\bibfnamefont {C.}~\bibnamefont {{\"U}nal}}, \bibinfo {author} {\bibfnamefont {F.}~\bibnamefont {Pacucci}}, \ and\ \bibinfo {author} {\bibfnamefont {A.}~\bibnamefont {Loeb}},\ }\href {\doibase 10.1088/1475-7516/2021/05/007} {\bibfield  {journal} {\bibinfo  {journal} {JCAP}\ }\textbf {\bibinfo {volume} {05}},\ \bibinfo {pages} {007} (\bibinfo {year} {2021})},\ \Eprint {http://arxiv.org/abs/2012.12790} {arXiv:2012.12790 [hep-ph]} \BibitemShut {NoStop}%
\bibitem [{\citenamefont {Hoof}\ \emph {et~al.}(2024)\citenamefont {Hoof}, \citenamefont {Marsh}, \citenamefont {Sisk-Reyn{\'e}s}, \citenamefont {Matthews},\ and\ \citenamefont {Reynolds}}]{Hoof:2024quk}%
  \BibitemOpen
  \bibfield  {author} {\bibinfo {author} {\bibfnamefont {S.}~\bibnamefont {Hoof}}, \bibinfo {author} {\bibfnamefont {D.~J.~E.}\ \bibnamefont {Marsh}}, \bibinfo {author} {\bibfnamefont {J.}~\bibnamefont {Sisk-Reyn{\'e}s}}, \bibinfo {author} {\bibfnamefont {J.~H.}\ \bibnamefont {Matthews}}, \ and\ \bibinfo {author} {\bibfnamefont {C.}~\bibnamefont {Reynolds}},\ }\href {\doibase 10.1093/mnras/staf1564} {\  (\bibinfo {year} {2024}),\ 10.1093/mnras/staf1564},\ \Eprint {http://arxiv.org/abs/2406.10337} {arXiv:2406.10337 [hep-ph]} \BibitemShut {NoStop}%
\bibitem [{\citenamefont {Witte}\ and\ \citenamefont {Mummery}(2025)}]{Witte:2024drg}%
  \BibitemOpen
  \bibfield  {author} {\bibinfo {author} {\bibfnamefont {S.~J.}\ \bibnamefont {Witte}}\ and\ \bibinfo {author} {\bibfnamefont {A.}~\bibnamefont {Mummery}},\ }\href {\doibase 10.1103/PhysRevD.111.083044} {\bibfield  {journal} {\bibinfo  {journal} {Phys. Rev. D}\ }\textbf {\bibinfo {volume} {111}},\ \bibinfo {pages} {083044} (\bibinfo {year} {2025})},\ \Eprint {http://arxiv.org/abs/2412.03655} {arXiv:2412.03655 [hep-ph]} \BibitemShut {NoStop}%
\bibitem [{\citenamefont {Arvanitaki}\ \emph {et~al.}(2024)\citenamefont {Arvanitaki}, \citenamefont {Madden},\ and\ \citenamefont {Van~Tilburg}}]{Arvanitaki:2021wjk}%
  \BibitemOpen
  \bibfield  {author} {\bibinfo {author} {\bibfnamefont {A.}~\bibnamefont {Arvanitaki}}, \bibinfo {author} {\bibfnamefont {A.}~\bibnamefont {Madden}}, \ and\ \bibinfo {author} {\bibfnamefont {K.}~\bibnamefont {Van~Tilburg}},\ }\href {\doibase 10.1103/PhysRevD.109.072009} {\bibfield  {journal} {\bibinfo  {journal} {Phys. Rev. D}\ }\textbf {\bibinfo {volume} {109}},\ \bibinfo {pages} {072009} (\bibinfo {year} {2024})},\ \Eprint {http://arxiv.org/abs/2112.11466} {arXiv:2112.11466 [hep-ph]} \BibitemShut {NoStop}%
\bibitem [{\citenamefont {Hook}(2018)}]{Hook:2018jle}%
  \BibitemOpen
  \bibfield  {author} {\bibinfo {author} {\bibfnamefont {A.}~\bibnamefont {Hook}},\ }\href {\doibase 10.1103/PhysRevLett.120.261802} {\bibfield  {journal} {\bibinfo  {journal} {Phys. Rev. Lett.}\ }\textbf {\bibinfo {volume} {120}},\ \bibinfo {pages} {261802} (\bibinfo {year} {2018})},\ \Eprint {http://arxiv.org/abs/1802.10093} {arXiv:1802.10093 [hep-ph]} \BibitemShut {NoStop}%
\bibitem [{\citenamefont {Di~Luzio}\ \emph {et~al.}(2021)\citenamefont {Di~Luzio}, \citenamefont {Gavela}, \citenamefont {Quilez},\ and\ \citenamefont {Ringwald}}]{DiLuzio:2021pxd}%
  \BibitemOpen
  \bibfield  {author} {\bibinfo {author} {\bibfnamefont {L.}~\bibnamefont {Di~Luzio}}, \bibinfo {author} {\bibfnamefont {B.}~\bibnamefont {Gavela}}, \bibinfo {author} {\bibfnamefont {P.}~\bibnamefont {Quilez}}, \ and\ \bibinfo {author} {\bibfnamefont {A.}~\bibnamefont {Ringwald}},\ }\href {\doibase 10.1007/JHEP05(2021)184} {\bibfield  {journal} {\bibinfo  {journal} {JHEP}\ }\textbf {\bibinfo {volume} {05}},\ \bibinfo {pages} {184} (\bibinfo {year} {2021})},\ \Eprint {http://arxiv.org/abs/2102.00012} {arXiv:2102.00012 [hep-ph]} \BibitemShut {NoStop}%
\bibitem [{\citenamefont {Co}\ \emph {et~al.}(2024)\citenamefont {Co}, \citenamefont {Gherghetta}, \citenamefont {Liu},\ and\ \citenamefont {Lyu}}]{Co:2024bme}%
  \BibitemOpen
  \bibfield  {author} {\bibinfo {author} {\bibfnamefont {R.~T.}\ \bibnamefont {Co}}, \bibinfo {author} {\bibfnamefont {T.}~\bibnamefont {Gherghetta}}, \bibinfo {author} {\bibfnamefont {Z.}~\bibnamefont {Liu}}, \ and\ \bibinfo {author} {\bibfnamefont {K.-F.}\ \bibnamefont {Lyu}},\ }\href {\doibase 10.1007/JHEP09(2024)145} {\bibfield  {journal} {\bibinfo  {journal} {JHEP}\ }\textbf {\bibinfo {volume} {09}},\ \bibinfo {pages} {145} (\bibinfo {year} {2024})},\ \Eprint {http://arxiv.org/abs/2407.12930} {arXiv:2407.12930 [hep-ph]} \BibitemShut {NoStop}%
\bibitem [{\citenamefont {Meng}(2016)}]{Meng2016}%
  \BibitemOpen
  \bibinfo {editor} {\bibfnamefont {J.}~\bibnamefont {Meng}},\ ed.,\ \href@noop {} {\emph {\bibinfo {title} {Relativistic Density Functional for Nuclear Structure}}},\ \bibinfo {series} {International Review of Nuclear Physics}, Vol.~\bibinfo {volume} {10}\ (\bibinfo  {publisher} {World Scientific, Singapore},\ \bibinfo {year} {2016})\BibitemShut {NoStop}%
\end{thebibliography}%

\clearpage
\onecolumngrid
\newpage

\setlength{\parindent}{15pt}
\setlength{\parskip}{1em}


\begin{center}
    {\Large \bfseries Supplemental Material for \\ ``Probing Axion via Mössbauer Spectroscopy''}
\end{center}

{{\bf A.} \it Data analysis strategy}

The ultra-narrow linewidth of the $^{109}$Ag Mössbauer transition imposes a critical constraint on the detector design. A detector that is too wide will cause a gravitational redshift broadening that exceeds the intrinsic linewidth. Conversely, an excessively narrow detector limits the photon count, which can yield insufficient statistics for peak reconstruction. While this trade-off makes a high photon count challenging, the persistent nature of the axion dark matter signal permits data collection over extended periods. Consequently, by leveraging long observation time, it becomes feasible to accumulate the necessary statistics despite the low instantaneous count rate.

With sufficient statistics, the precision of the peak location measurement is ultimately limited by systematic uncertainties. Through careful calibration and experimental tuning, the potential systematic uncertainties can be mitigated. It is reasonable to assume that the total systematic uncertainty can be reduced  to a level comparable to the frequency broadening due to the detector's finite size. Our goal is to balance the statistical and systematic uncertainties and to achieve the best sensitivity to the axion search.

The axion dark matter induces an oscillatory resonance peak location along the vertical direction
\begin{equation}
    Z_p(t) =  A \cos(\omega t +\phi).
\end{equation}
Here $A$ is the oscillation amplitude, which is related to the PQ symmetry breaking scale $f_a$. The $\omega$ is determined by the axion mass ($\omega \simeq 2 m_a$) and $\phi$ is a phase.

Assuming the experiment collects $N_{\rm tot}$ data points over a total time $T_{\rm tot}$. For an axion which induces a signal with oscillation period $T_a=2\pi/\omega$, the coherence time is $T_{\rm coh}\sim 10^6 T_a$. In order to maximize the sensitivity, we propose the following signal extraction strategy:
\begin{itemize}
\item (1) Segment by Coherence Time: 

Divide the total dataset 
$D$ into $N_c=T_{\rm tot}/T_{\rm coh}$ segments, labeled as $D_i (i=1,...,N_c)$.

\item (2) Segment by Oscillation Period: 

Within each coherent segment $D_i$, subdivide the data into $10^6$ smaller segments, $D_{i,\alpha} (\alpha=1,...,10^6)$, each corresponding to one oscillation period $T_a$.

\item (3) Define Phase Bins: 

Further divide each oscillation period into $n_{\rm ph}$ phase bins, resulting in data subsets $D_{i,\alpha,m}$, where $m=1,...,n_{\rm ph}$ represents the phase. Each bin corresponds to the time duration of $\delta T = T_a/n_{\rm ph}$.

\item (4) Coherent Stacking: 

For a fixed coherence segment $i$
and a fixed phase bin $m$, sum the data $D_{i,\alpha,m}$ over all periods $\alpha$. This step ensures that the axion signal, which shares a common phase across different periods, adds constructively.

After this stacking procedure, each coherent segment $D_i$ is reduced to $n_{\rm ph}$ phase bins. The number of events in each phase bin is approximately:
\begin{equation}
    N_{\rm ph} \simeq \frac{N_{\rm tot}}{n_{\rm ph}}\frac{T_{\rm coh}}{T_{\rm tot}} \ .
\end{equation}
\item (5) Optimization:

To optimize the search sensitivity, we balance statistical and systematic uncertainties. The statistical uncertainty for each phase bin is  $1/\sqrt{N_{\rm ph}}$, assuming $N_{\rm ph}\gtrsim 10$.  

Along the vertical direction, the photon flux is expected to follow a Lorentzian distribution, 
\begin{equation}
    1 - f_S \, \epsilon \cdot 
\frac{\Gamma_{0}^2}{
\left[ g(Z - Z_p) E_\gamma \right]^2 + \Gamma_{0}^2} \,
\end{equation}
where $Z_p$ is the peak location and $E_\gamma$ is photon energy at the resonance, i.e. 88 keV for $^{109}$Ag.

For each phase bin $m$, a $\chi^2$ fitting can be performed to find the peak location. 
Statistical fluctuations lead to an uncertainty in the fitted value of $Z_p$ as
 ~\cite{Gao:2023ggo}
\begin{equation}
    \dfrac{\delta Z_p}{\Delta Z} = \dfrac{\xi(\epsilon f_S)}{\sqrt{N_{\rm ph}}}, 
\end{equation}
where $\xi(x) =  -0.17 + 0.16 x^{-1} + 0.014 x^{-2}$.

The optimal number of phase bins, $n_{\rm ph}$, is determined by requiring that the statistical uncertainty to be comparable to the frequency resolution induced by the detector's width, $\Delta Z$, i.e. $\dfrac{\delta Z_p}{\Delta Z}\sim 1$. 

\item (5) Combining coherence time segments:

At last, by combining $N_c$ number of the coherence time segments, we obtain the precision of the oscillatory amplitude $A$ as
\begin{equation}
     \delta A \simeq \dfrac{\sqrt{2} \Delta Z}{\sqrt{n_{\rm ph}} \sqrt{N_c}}
\end{equation}

\end{itemize}

Several subtleties must be addressed when implementing the proposed analysis strategy. First, the finite size of the source itself imposes a limit on the achievable frequency resolution. To ensure that the emitted photon flux can be treated as originating from a single effective height, the vertical extent of the source must be smaller than the width of the detector.

Second, the integration time for each phase bin, denoted as $\delta T$ in Step (3), must exceed the detector's intrinsic time resolution, $t_{\rm res}$. This requirement, i.e. $\delta T = T_a/n_{\rm ph} \geq t_{\rm res}$,  constrains the maximum axion mass that can be probed, leading to the condition
$m_a \leq 2\pi/(n_{\rm ph} t_{\rm res})$.

Furthermore, to unambiguously reconstruct the periodic signal, the sampling must satisfy the Nyquist-Shannon criterion. This requires a sampling frequency of at least twice the signal frequency. 
Here we conservatively impose $n_{\rm ph} \geq 4$ to be safe. Consequently, for a detector with a time resolution of $t_{\rm res} = 1$ ns, the experiment is sensitive to axion masses up to approximately $m_a \lesssim 1\ \mu\text{eV}$.

Conversely, the proposed analysis strategy becomes ineffective if the axion mass is too small, specifically when the coherence time exceeds the observation time, approximately one year. In this regime, one can no longer segment the data by coherence time. Nevertheless, a search remains viable by applying the phase-binning procedure described previously to the entire dataset, thereby coherently stacking the signal across the full observation period.

Finally, to ensure physical consistency within the framework of a QCD axion, we impose the requirement that the axion decay constant $f_a$
remains sub-Planckian, i.e. $f_a < M_{pl}$. 

{\bf B.}\textit{Details about the RDFT calculations}

The starting point of the RDFT is the following relativistic Lagrangian density,
\begin{equation}\label{eq:lagrangian}
  \begin{split}
    \mathcal{L} &= \bar{\psi}(i\gamma_\mu\partial^\mu - m)\psi\\
    &-\frac{1}{2}\alpha_S(\bar{\psi}\psi)(\bar{\psi}\psi) - \frac{1}{2}\alpha_V(\bar{\psi}\gamma_\mu\psi)(\bar{\psi}\gamma^\mu\psi) \\
    &-\frac{1}{2}\alpha_{TV}(\bar{\psi}\vec{\tau}\gamma_\mu\psi)(\bar{\psi}\vec{\tau}\gamma^\mu\psi)\\
    & -\frac{1}{3}\beta_S(\bar{\psi}\psi)^3 - \frac{1}{4}\gamma_S(\bar{\psi}\psi)^4 - \frac{1}{4}\gamma_V[(\bar{\psi}\gamma_\mu\psi)(\bar{\psi}\gamma^\mu\psi)]^2\\
    &-\frac{1}{2}\delta_S\partial_\nu(\bar{\psi}\psi)\partial^\nu(\bar{\psi}\psi) - \frac{1}{2}\delta_V\partial_\nu(\bar{\psi}\gamma_\mu\psi)\partial^\nu(\bar{\psi}\gamma^\mu\psi)\\
    &-\frac{1}{2}\delta_{TV}\partial_\nu(\bar{\psi}\vec{\tau}\gamma_\mu\psi)\partial^\nu(\bar{\psi}\vec{\tau}\gamma^\mu\psi)\\
    &-\frac{1}{4}F^{\mu\nu}F_{\mu\nu} - e\frac{1-\tau_3}{2}\bar{\psi}\gamma^\mu\psi A_\mu.
  \end{split}
\end{equation}
It includes nine coupling constants: $\alpha_S$, $\alpha_V$, $\alpha_{TV}$, $\beta_S$, $\gamma_S$, $\gamma_V$, $\delta_S$, $\delta_V$, and $\Delta_{TV}$, where the subscripts $S$, $V$, and $TV$ denote the scalar, vector, and isovector couplings, respectively.
Based on this Lagrangian density, the nuclear Hamiltonian can be derived via the Legendre transformation.
Within the meanfield approximation, the nuclear trial wavefunction is represented as a Slater determinant $|\Phi\rangle = \prod_i^A a^\dagger_i|0\rangle$, with $a_i^\dagger$ being the creation operator for the single nucleonic state $\psi_i$.
By expanding the nucleonic field $\psi$ and $\bar{\psi}$ in terms of the basis states $\psi_i$, the Hamiltonian can be rewritten in a second-quantized form.
Its expectation value with respect to $|\Phi\rangle$ yields the total energy of the nuclear system, an energy functional that depends on nuclear one-body densities and currents.
The variation of the total energy with respect to the single-particle state $\psi_i$ results in the relativistic Kohn-Sham equation.

For open-shell nuclei, such as the $^{109}$Ag studied in this work, pairing correlations must be taken into account.
A unified treatment of both meanfield and pairing correlations is achieved by extending the relativistic Kohn-Sham equation to the following RHB equation,
\begin{equation}\label{eq:RHB}
  \left(\begin{array}{cc}
    \hat{h}_D-\lambda& \hat{\Delta}\\
    -\hat{\Delta}^\ast & -\hat{h}_D^\ast + \lambda
  \end{array}\right)
  \left(\begin{array}{c}
    U_k\\
    V_k
  \end{array}\right) = E_k
  \left(\begin{array}{c}
    U_k\\
    V_k
  \end{array}\right),
\end{equation}
where $\hat{h}_D$ and $\hat{\Delta}$ denote the single-nucleon Hamiltonian and the pairing field, respectively.
By iteratively solving the RHB equation, both the nuclear wavefunction and the total energy can be self-consistently obtained.
For further details, refer to Ref.~\cite{Meng2016}.

\begin{table}[htbp!]
  \centering
  \caption{Total energies of the ground state ($E_{1/2-}$) and the first excited state ($E_{7/2+}$) in $^{109}$Ag calculated by the RDFT based on the PC-PK1 density functional.
  The contributions of various coupling channels to the total energies as well as the energy difference $\Delta E = E_{7/2+} - E_{1/2-}$ are also presented.
  Here, $E_{\mathrm{kin}}$, $E_{\mathrm{cou}}$, $E_{\mathrm{pair}}$, and $E_{\mathrm{cm}}$ denote the kinetic energy, Coulomb energy, pairing energy, and the center-of-mass correction energy, respectively.}
  \begin{tabular}{cccc}
  \hline\hline
    Channels   & $1/2^-$ (MeV) & $7/2^+$ (MeV) & $\Delta E $ (MeV) \\
  \hline 
    $E_{\alpha_S}$      &  -17623.997                &  -17510.543              &  +112.997    \\
    $E_{\alpha_V}$      &  +13065.872                &  +12992.183              &  -73.689     \\
    $E_{\alpha_{TV}}$   &  +19.427                   &  +19.152                 &  -0.273      \\
    $E_{\beta_S}$       &  +2539.210                 &  +2510.932               &  -28.278     \\
    $E_{\gamma_S}$      &  -870.667                  &  -857.668                &  +12.998     \\
    $E_{\gamma_V}$      &  -100.616                  &  -99.473                 &  +1.143      \\
    $E_{\delta_S}$      &  +46.431                   &  +44.635                 &  -1.796      \\
    $E_{\delta_V}$      &  +197.220                  &  +189.950                &  -7.269      \\
    $E_{\delta_{TV}}$   &  +2.166                    &  +2.121                  &  -0.045      \\ 
    $E_{\mathrm{kin}}$  &  +1477.263                 &  +1461.798               &  -15.465     \\
    $E_{\mathrm{cou}}$  &  +334.000                  &  +334.475                &  +0.475      \\
    $E_{\mathrm{pair}}$ &  -8.908                    &  -10.347                 &  -1.439      \\
    $E_{\mathrm{cm}}$   &  -6.673                    &  -6.516                  &  +0.158      \\
    $E_{\mathrm{tot}}$  &  -928.814                  &  -929.298                &  -0.483      \\ 
  \hline\hline 
  \end{tabular}\label{Tab:Diff-E-total}
\end{table}

Based on the RDFT, we calculate the total energies of the ground state $(J^\pi = 1/2^-)$ and the excited state $(J^\pi = 7/2^+)$ in $^{109}$Ag.
The obtained results, along with the contributions from individual coupling channels defined in Eq.~\eqref{eq:lagrangian}, are presented in Tab~.\ref{Tab:Diff-E-total}.
As shown in the last column, the small total energy difference $\Delta E$ arises from the cancellation among the contributions of individual coupling channels.
The most significant cancellation occurs between the $\alpha_S$ and $\alpha_V$ channels, whose contributions to $\Delta E$ are as large as $+112.997$ MeV and $-73.689$ MeV, respectively.

\end{document}